\documentclass[trackchanges,twocolumn]{aastex}

\let\tablenum\relax

\usepackage{natbib,epsfig,siunitx,url,graphicx,amsmath,footnote,float,xcolor,cases}
\begin{document}

\submitted{ApJ; accepted}

\title{Habitable planet formation around low-mass stars: Rapid accretion, rapid debris removal and the essential contribution of external giants}

\author{Matthew S. Clement\altaffilmark{1}, Elisa V. Quintana\altaffilmark{2} \& Billy L. Quarles\altaffilmark{3,4}}

\altaffiltext{1}{Earth and Planets Laboratory, Carnegie Institution for Science, 5241 Broad Branch Road, NW, Washington, DC 20015, USA}
\altaffiltext{2}{NASA Goddard Space Flight Center, Greenbelt, MD 20771, USA}
\altaffiltext{3}{Center for Relativistic Astrophysics, School of Physics, Georgia Institute of Technology, Atlanta, GA 30332 USA}
\altaffiltext{4}{Department of Physics, Astronomy, Geosciences and Engineering Technology, Valdosta State University, Valdosta GA, 31698, USA}
\altaffiltext{*}{corresponding author email: mclement@carnegiescience.edu}

\begin{abstract}

In recent years a paradigm shift has occurred in exoplanet science, wherein low-mass stars are increasingly viewed as a foundational pillar of the search for potentially habitable worlds in the solar neighborhood. However, the formation processes of this rapidly accumulating sample of planet systems are still poorly understood. Moreover, it is unclear whether tenuous primordial atmospheres around these Earth-analogs could have survived the intense epoch of heightened stellar activity that is typical for low-mass stars.  We present new simulations of in-situ planet formation across the M-dwarf mass spectrum, and derive leftover debris populations of small bodies that might source delayed volatile delivery.  We then follow the evolution of this debris with high-resolution models of real systems of habitable zone planets around low-mass stars such as TRAPPIST-1, Proxima Centauri and TOI-700. While debris in the radial vicinity of the habitable zone planets is removed rapidly, thus making delayed volatile delivery highly unlikely, we find that material ubiquitously scattered into an exo-asteroid belt region during the planet formation process represents a potentially lucrative reservoir of icy small bodies. Thus, the presence of external $\sim$Neptune-Saturn mass planets capable of dynamically perturbing these asteroids would be a sign that habitable zone worlds around low-mass stars might have avoided complete desiccation.  However, we also find that such giant planets significantly limit the efficiency of asteroidal implantation during the planet formation process.  In the coming decade, long-baseline radial velocity studies and Roman Space Telescope microlensing observations will undoubtedly further constrain this process.

\end{abstract}

\section{Introduction}

While the habitable zones (HZs; circumstellar regions where liquid water might exist on rocky surfaces) of Sun-like stars disproportionately host planets in the Super-Earth-mass regime \citep{chiang13,zhu18}, exoplanet demographics \citep{dressing13,dressing15,gaidos16} tend to support theoretical predictions \citep[e.g.:][]{ogihara09,miguel11} that low-mass stars more frequently harbor $\sim$Earth-mass planets at distances that might be hospitable to the emergence of life.  In addition to small rocky planets seemingly representing a common end-state of the general planet formation process around M-Dwarfs, such stars comprise the most common stellar class in the galaxy \citep{henry06}, and the systems' large planet-star size ratios \citep[and correspondingly deep transits:][]{dewit16,pidhorodetska21} make them ideal candidates for atmospheric characterization efforts.  Among others, the discovery of potentially habitable planetary systems such as those around Kepler 186 \citep{quintana14}, TRAPPIST-1 \citep{gillon16} and Proxima Centauri \citep{proxima} have recently invigorated attempts to develop more robust models for the formation of habitable worlds around small stars \citep[e.g.:][]{ormel17,papaloizou17,grimm18,schoonenberg19}

While the prospect of the Transiting Exoplanet Survey Satellite \citep[TESS:][]{ricker15,barclay18} mission detecting a sizable sample of M-Dwarf-hosted transiting planets that are Earth-like in mass, radius and incident irradiation is compelling, such planets likely face a myriad of challenges beyond those typically considered within the paradigm of habitability as understood in the solar system.  Strong tides \citep{bolmont11}, enhanced stellar activity \citep{hawley14}, and prolonged periods of exceedingly high XUV-fluxes during the small stars' long cooling timescales en-route to their arrival on the main sequence \citep[in excess of 1 Gyr for stars like TRAPPIST-1:][]{baraffe15,bolmont17} represent significant roadblocks to the emergence and sustainability of life in these environments \citep[e.g.:][]{tian15}.  More specifically, heightened XUV irradiation over such timescales is likely capable of desiccating the volatile inventories of would-be Earth-analogs by evaporating oceans and photodissociating the resultant vapor.  Through this process, low-mass hydrogen atoms become vulnerable to atmospheric escape (as would the constituencies of primordial H-He atmospheres around similar planets), and other molecules are reprocessed (e.g.: CO $+$ OH $\rightarrow$ CO$_{2}$ $+$ H).  If this is the case, the potential habitability of these planets hinges on their ability to develop secondary atmospheres; either through the delivery of icy small bodies \citep[e.g.:][]{kral18,dencs19,ray21} or by the outgassing of initially enhanced internal volatile contents \citep{bolmont17,bourrier17}

In spite of considerable effort, the TRAPPIST-1 planets continue to befuddle atmospheric detection campaigns utilizing transmission spectroscopy \citep{dewit16,dewit18,gillon17,luger17,wakeford19}.  Given the degeneracies between plausible planetary compositions in mass-radius space \citep{zeng13,baraffe13}, computational models of planet formation in these systems remain an essential tool for probing their potential internal and atmospheric structures \citep[e.g.:][]{papaloizou17,quarles17,grimm18,dencs19,coleman19,djovsovic20,hori20}.  From a theoretical standpoint, considerable focus has been placed on replicating the specific TRAPPIST-1 system \citep[e.g.:][]{coleman19,schoonenberg19,huang21}.  However, the more generic process of rocky planet formation across the broad M-Dwarf mass spectrum \citep[$\sim$0.07-0.6 $M_{\odot}$:][]{auddy16} is still poorly understood \citep[e.g.:][]{ray07_mdwarf,lissauer07,miguel11,dugaro16,matsumoto20,miguel20}.  While the resonant architecture of multi-planet systems around low-mass stars like TRAPPIST-1 \citep{luger17_res} and TOI-178 \citep[a $\sim$0.6 $M_{\odot}$ star:][]{leleu21_res} strongly suggest that convergent migration played an important role in their formation \citep{tamayo17,izidoro19}, it is unclear whether this process is ubiquitous.  Moreover, the overall significance and role of pebble accretion \citep[e.g.:][]{johansen10,morbidelli12,levison15,ida16,chambers16}, planetesimal accretion \citep[e.g.:][]{koko_ida_96,chambers06,clement20_psj,woo21} and late-stage giant impacts \citep[e.g.:][]{chambers98,ray09a,quintana15,walsh19} in the formation of the solar systems' terrestrial planets is still debated, thus complicating the mapping of the problem to exoplanets.  

While systems of short-period rocky planets strongly resemble scaled-up versions of the solar system's giant planet satellite systems \citep{kane13,miguel20}, it is likely that the varied configurations of moons around Jupiter, Saturn, Uranus and Neptune also evince a complex and diverse range of formation and evolutionary histories \citep[e.g.:][]{canup02,nesvory07,cuk16,batygin20,madeira21}.  Nevertheless, it is generally accepted that short-period, rocky planets around low-mass stars must form through one of two possible avenues: (1) in-situ via planetesimal impacts with potential contributions from pebbles \citep{ray07_mdwarf,hansen15}, or (2) further out in the primordial nebular disk with more significant contributions from pebbles and subsequent implantation in the HZ via Type I migration \citep{ogihara09,ormel17}.

In this paper, we turn our attention to the question of volatile acquisition, retention and replenishment in rocky planets around low-mass stars.  Specifically, we are interested in the potential of small bodies such as leftover planetesimals and icy asteroids (i.e.: exo-asteroid belts) to deliver volatile-rich material capable of reconstituting the atmospheres of desiccated planets long after their formation \citep[e.g.:][]{kral18,dencs19,djovsovic20,ray21}.  While the smallest objects typically considered in planet formation simulations have $D\simeq$ 1,000 km, cratering records on solar system bodies \citep[e.g.:][]{minton19} indicates that the actual sizes of these objects were more similar to those of main belt asteroids ($\sim$1-100 km).  As leftover debris is a natural outcome of the in-situ formation of the solar system's terrestrial planets \citep{ray13,clement18_frag} that likely played a critical role in the evolution of Earth's atmosphere \citep{sinclair20} and potentially implanted a significant number of main belt asteroids \citep{izidoro16,ray17sci} that continue to feed today's near-Earth asteroid population, it naturally follows to examine these bombardment sources within the in-situ model of planet formation around low-mass stars.  To better understand these processes, we first perform a suite of numerical simulations to study the accretion of terrestrial planets in systems spanning the complete M-Dwarf mass spectrum.  Motivated by recent planets detected in high-precession radial velocity (RV) campaigns \citep{tuomi14,feng20a,feng20b} and microlensing surveys \citep{street16,ranc19,jung19}, we incorporate a batch of models that consider perturbations from an external, $\sim$Neptune-mass planet on the growth of the interior planets.  We then model the long-term stability and evolution of the resultant debris populations from these accretion simulations within several multi-planet systems of detected planets around low mass stars \citep{gillon16,proxima,gilbert20}.

\section{Methods}

\subsection{Planet Formation Simulations}
\label{sect:meth_form}

Our N-body planet formation simulations utilize the well-established \textit{Mercury6} Hybrid integration package \citep{chambers99}.  While the resolution of our simplified numerical models is admittedly low compared to many contemporary models of planet formation \citep[e.g.:][]{morishma10,carter15,woo21}, this is acceptable for our purposes as the goal of our study is broadly investigate planet formation around M-Dwarfs and follow the dynamical evolution of debris \citep[i.e.: collisional fragments and leftover planetesimals:][]{genda12,lands12,quintana15,clement18_frag} generated and stranded during the ultimate epoch of giant impacts in these systems.  Thus, we limit the number of simulation particles employed in our initial simulation set and do not include a pebble accretion model to minimize compute time and perform a broad parameter space sweep before switching to higher-resolution models in our investigation of post-formation bombardment (section \ref{sect:meth_bomb}).  Moreover, as the initial conditions for planet formation around low-mass stars remain poorly understood and largely unconstrained, we begin our investigation by deviating only slightly from the accepted paradigm of rocky planet accretion in the solar system \citep[e.g.:][]{chambers98,chambers01,hansen09}.  An important limitation of this methodology is the absence of a gas disk model in our simulations.  If short-period planets around M-Dwarfs form in-situ, their growth is sufficiently rapid that it must have coincided with the gas disk phase \citep[unless planetesimal formation is delayed for some reason][see section \ref{sect:discuss} for a more detailed discussion of these caveats:]{ogihara15}.  We plan to expound upon the results of this preliminary study in future work with more detailed models and artificial gas disk treatments \citep[e.g.:][]{clement20_psj} specifically tuned to form specific systems of M-Dwarf-hosted planets \citep[e.g.:][]{ormel17,coleman19,schoonenberg19}.

Our simulations study rocky planet formation across the M-Dwarf mass spectrum.  Specifically, we consider stellar masses of 0.1-0.6 $M_{\odot}$ in 0.1 $M_{\odot}$ increments (36 total simulations per mass).  Each planet formation simulation follows the collisional evolution of a disk of planet-forming material distributed between 0.01 and 0.5 au, with a surface density profile that falls off radially as $r^{-3/2}$ \citep[e.g.:][]{birnstiel12}.  The inner edge of our disks are loosely based off the magnetic truncation radius \citep[e.g.:][]{frank92,ormel17}: 

\begin{equation}
	a_{in} = 0.01 \bigg( \frac{B_{*}}{180 G} \bigg)^{4/7} \bigg( \frac{R_{*}}{0.5 R_{\odot}} \bigg)^{12/7} \bigg( \frac{M_{*}}{0.1 M_{\odot}} \bigg)^{-2/7}
\end{equation}As the size of the magnetospheric cavity only varies mildly with stellar mass, we utilize the same interior disk boundary of 0.01.0 au for each stellar mass investigated.  In light of recent detections of $\sim$10-100 $M_{\oplus}$ planets with semi-major axes of $\sim$0.5-3.0 au around M Dwarfs \citep[e.g.: GJ 9066b, HD 147379b, GJ 229b, GJ 433c, HIP 38594c, GJ 9066c:][]{tuomi14,carmens18,feng20a,feng20b}, we place a Neptune-mass giant planet at 1.0 au in half of our simulations.  This positions our outer terrestrial disk boundary (0.5 au) approximately 8-12 Hill spheres away from the external perturber \citep{chambers96}. 

Table \ref{table:ics} summarizes the initial conditions for our various simulation sets.  Each system is integrated for 20 Myr \citep[which we find to be a reasonable amount of time for the planetary building blocks to coalesce into relatively stable configuration of $\sim$4-8 planets:][]{ray07_mdwarf} using a time-step equal to 5$\%$ the orbital period at 0.01.0 au \citep{chambers01}.  Particles are merged with the central body at perihelia passages less than 0.002 au, and considered ejected at heliocentric distances of 3 au.  Our calculations also include artificial forces to account for the effects of general relativity \citep{saha92}.  Semi-major axes of simulation particles are selected to achieve our $\Sigma_{solid} \propto r^{-3/2}$ surface density profile.  Eccentricities and inclinations are initialized by sampling near-circular Rayleigh distributions ($\sigma_{e}=$ 0.02; $\sigma_{i}=$ 0.2$\degr$), while the remaining orbital elements are drawn from uniform distributions of angles.  In all cases, planetary embryos \citep[presumably formed by runaway accretion of planetesimals:][]{koko_ida_96} are assigned equal masses, and treated as fully self-gravitating by our code i.e.: they gravitationally interact with all objects in the simulation).  A swarm of equal-mass planetesimals \citep[e.g.:][]{ray06,obrien06} constituting half the total disk mass are distributed in all but 24 or our simulations (table \ref{table:ics}).  Moreover, the planetesimals are treated as semi-active particles (they feel the gravitational forces of the embryos but not one another).  In all of these initial planet formation simulations collisions are treated as perfect mergers (we test the effects of fragmentation in a follow-on suite of simulations, see section \ref{sect:frag})  The purpose of our embryo-only simulations is to crudely approximate in-situ formation in a scenario where inward migrating proto-planets pile-up and accrete at the disk's inner edge \citep{miguel20}.  Moreover, these simulations confirm (see section \ref{sect:bomb}) that systems of large proto-planets undergoing giant impacts still strand debris in the form of collisional fragments generated in the ultimate series of massive mergers.

Estimating an initial disk mass for in-situ planet formation around low-mass stars is complicated by many large uncertainties in the properties of the stars' natal disks, as well as the fact that the complete range of fully evolved dynamical structures is relatively unconstrained.  Debris disks have not been definitively detected around such stars, thus making it difficult to place meaningful observational constraints on their planet formation environments \citep[e.g.:][]{flaherty19}.   While early investigations argued that disk masses likely scaled directly with stellar mass, thus making Earth-mass planets in the HZ \citep{ray07_mdwarf,lissauer07} and giant planet formation \citep{laughlin04} unlikely around the smallest stars, these assumptions have since been contradicted by both Kepler's planet yield \citep{dressing13,gaidos16} and high-precession radial velocity studies \citep{feng20a,feng20b}.  Indeed, \citet{dressing15} estimate that M-Dwarfs host an average of 2.5 planets with orbital periods shorter than 100 days and 1.0 $<R<$ 4.0 $R_{\oplus}$.  Thus, modern empirical estimates of the minimum mass extra-solar nebula \citep{chiang13} around M-Dwarfs argue that planets around such stars might form within massive disks possessing upwards of 5.0 $M_{\oplus}$ in solids interior to 0.5 au \citep[e.g.:][]{gaidos17,carmens21_mmsn}.  To maintain the focus of our investigation on the growth of Earth-mass planets in the HZs of low-mass stars, we test two different total terrestrial disk masses (3.0 and 6.0 $M_{\oplus}$) that are loosely derived from the observationally inferred minimum mass extra-solar nebula \citep{chiang13,gaidos17,carmens21_mmsn}.

\begin{table*}
\centering
\begin{tabular}{c c c c c c c}
\hline
$M_{*}$ ($M_{\odot}$) & $M_{disk}$ ($M_{\oplus}$) & Neptune $@$ 1.0 au & $N_{emb}$ & $N_{pln}$ & $N_{sim}$/$M_{*}$ & $N_{sim}$ \\
\hline
0.1,0.2,0.3,0.4,0.5,0.6 & 3.0 & Y & 20 & 200 & 8 & 48 \\
 & 3.0 & N & 20 & 200 & 8 & 48 \\
 & 6.0 & Y & 20 & 200 & 8 & 48\\
 & 6.0 & N & 20 & 200 & 8 & 48 \\
 & 3.0 & Y & 200 & 0 & 1 & 6 \\
 & 3.0 & N & 200 & 0 & 1 & 6 \\
 & 6.0 & Y & 200 & 0 & 1 & 6 \\
 & 6.0 & N & 200 & 0 & 1 & 6 \\
\hline
\end{tabular}
\caption{Summary of initial conditions for our various simulation sets.  The columns are as follows: (1) The various stellar masses investigated in solar units (note that we perform 36 total simulations for each stellar mass), (2) the total mass of terrestrial forming material in Earth units, (3) whether a Neptune-mass giant planet is placed at 1.0 au, (3) the total number of equal-mass, fully self-gravitating embryos comprising half the total disk mass (or all of the mass in cases where planetesimals are not utilized), (4) the total number of equal-mass, semi-interacting planetesimals, (7) the total number of simulations per stellar mass for each combination of varied parameters, and (8) the total number of simulations described by the row.}
\label{table:ics}
\end{table*}

Table \ref{table:stars} tabulates a number of presumed properties of the host stars in our numerical models.  Of particular importance, we derive the optimistic and conservative circumstellar HZs for each system utilizing the relationships presented in \citet{hz}.  These radial ranges are utilized for defining Earth-analogs and potentially habitable planets in our subsequent analyses (section \ref{sect:water}).  We also compute the approximate location of the water-ice snowline ($a_{snow}$) in each system \citep[loosely defined here as the location in the disk where T $=$ 170 K; $\sim$2.7 au in the solar nebula:][]{hayashi81}, and report the respective semi-major axes in table \ref{table:stars}.  Objects originating exterior to the snowline are assumed to begin with water-ice contents of 10$\%$ their total mass, those with semi-major axes within 20$\%$ of the inside of this boundary are initialized with 0.1$\%$ of their mass in the form of water, and those beginning the simulation inside of the 0.1$\%$ region are assigned values of 0.001$\%$.  This is motivated by the inferred compositions of main belt asteroids and solar system chondrites \citep[see][for recent reviews]{morb12,obrien18,meech19}.  Since the snowline's location is dependent on stellar luminosity \citep[e.g.:][]{ida04}, it should be noted that its precise orientation likely evolves substantially in M-Dwarfs \citep[e.g.:][]{kennedy08} that experience significant pre-Main Sequence cooling \citep{baraffe15}: 

\begin{equation}
	T = 280 \bigg( \frac{a}{1.0 au} \bigg) ^{-1/2} \bigg( \frac{L_{*}}{L_{\odot}} \bigg) ^{1/4} K
	\label{eqn:ice}
\end{equation}Thus, our calculations might over-estimate the water contents of fully accreted planets.  However, if the majority of our large embryos and planetesimals formed further out in the disk and were subsequently transported inwards via Type I migration \citep{tanaka04}, they might possess enhanced volatile contents relative to our assumptions.  Nevertheless, while it may be difficult to define a physically motivated initial compositional gradient for our terrestrial disks, structuring our analysis in this manner allows us to clearly compare the effects of mixing in our simulations considering different stellar masses and giant planet models.

\begin{table*}
\centering
\begin{tabular}{c c c c c c}
\hline
$M_{star}$ ($M_{\odot}$) & Spectral Class & approx. T$_{eff}$ (K) & HZ$_{consv.}$ (au) & HZ$_{opt}$ (au) & $a_{snow}$ (au) \\
\hline
0.1 & M6V & 2600 & .031-.064 & .025-.067 & .081 \\
0.2 & M4V & 3100 & .078-.16 & .061-.17 & .20  \\
0.3 & M3V & 3200 & .12-.23 & .091-.24 & .28 \\
0.4 & M2V & 3350 & .14-.28 & .11-.30  & .37 \\
0.5 & M1V & 3600 & .19-.37 & .15-.39 & .5 \\
0.6 & M0V & 3800 & .28-.53 & .22-.56 & $>$.5 \\
\hline
\end{tabular}
\caption{Summary of assumed stellar properties for the various analyses presented in this manuscript.  The columns are as follows: (1) the stellar mass investigated, (2) the corresponding main sequence spectral type, (3-4) the conservative and optimistic HZs as defined in \citep{hz,hz2}, and (5) the location of the water-ice line \citep[calculated via equation \ref{eqn:ice}, e.g.:][]{ida04}.}
\label{table:stars}
\end{table*}

\subsection{High-resolution bombardment models}
\label{sect:meth_bomb}

\subsubsection{Systems of Interest}

For the purposes of our current investigation, we are chiefly interested in the total masses and orbital distributions of planetesimals and collisional fragments that would be generated in (or stranded by) the ultimate epoch of giant impacts that complete the formation of our systems.  The second part of our analysis involves studying the dynamical evolution of this debris in three multi-planet, M-dwarf hosted systems \citep[TRAPPIST-1, Proxima Centauri and TOI-700:][]{gillon16,proxima,gilbert20}.  In all cases we utilize the current published orbits and masses for the systems' planets as inputs for our numerical simulations \citep{proxima,gillon17,trappist_mass,proxima_c,proxima_c_mass,proxima_new,trappist_new}.  TRAPPIST-1 is an 0.0898 $M_{\odot}$ M8V dwarf \citep{trappist_mass} with three planets in the HZ (e, f and g).  Proxima Centauri is an 0.12 $M_{\odot}$ M5.5V dwarf \citep{proxima_mass} with the $\sim$1.6 $M_{\oplus}$ planet Proxima Centauri b situated in the HZ. TOI-700 is an 0.41 $M_{\odot}$ M2V dwarf \citep{gilbert20} hosting three planets, the last of which (d) is potentially habitable and has a mass of $\sim$1.72 $M_{\oplus}$ \citep{rodriguez20}.  In the case of Proxima Centauri, we elect to include the suspected small, $\sim$0.3 $M_{\oplus}$ interior planet reported by \citet{proxima_new}.  We chose these systems because they encompass a range of stellar masses and each possess an $\sim$Earth-sized, potentially rocky planet in the HZ.

\subsubsection{Generating Fragment Populations}
\label{sect:frag}

\begin{figure*}
	\centering
	\includegraphics[width=.49\textwidth]{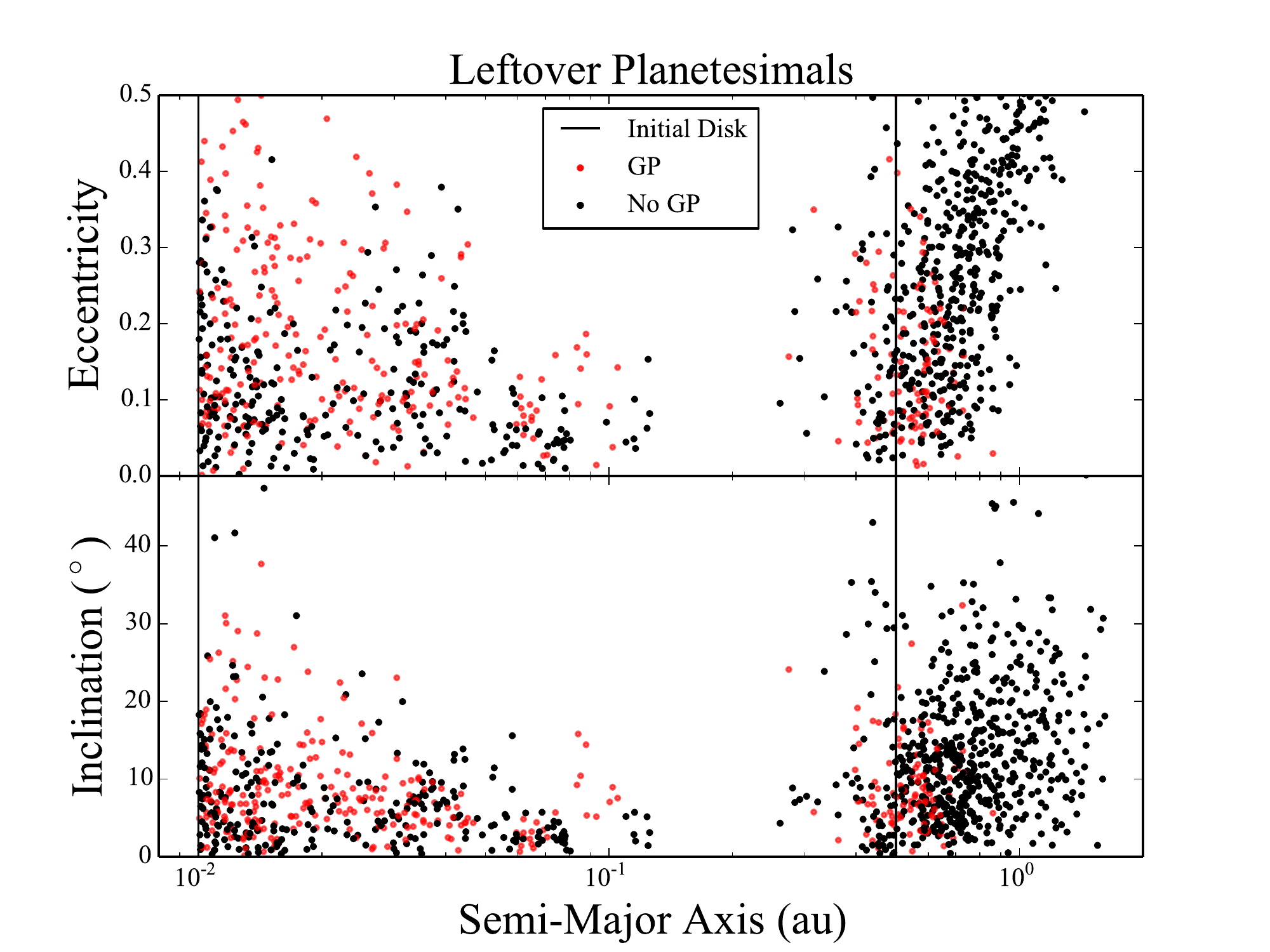}
	\includegraphics[width=.49\textwidth]{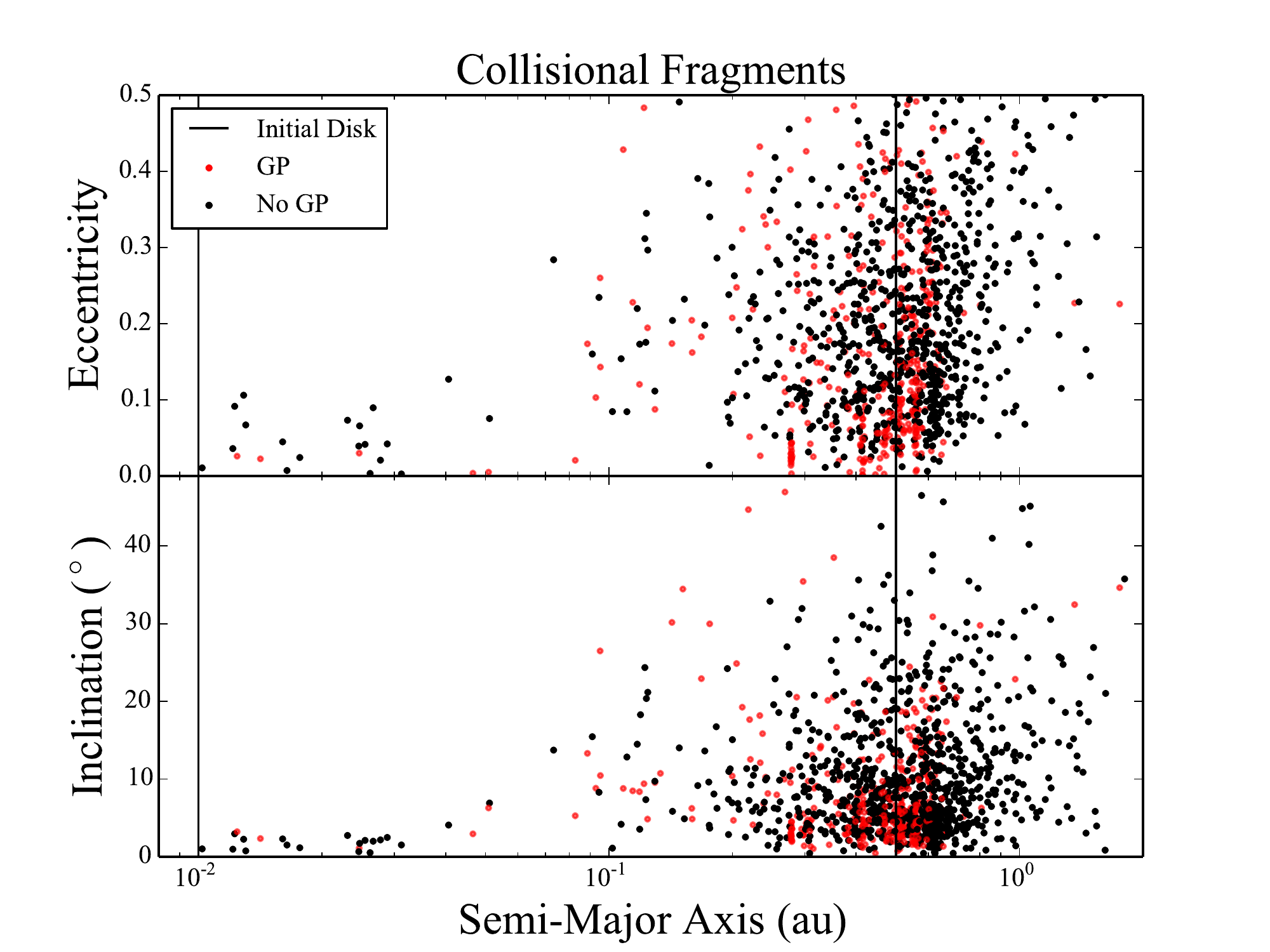}
	\caption{Left panel: Eccentricity and Inclination distributions of 1375 remnant planetesimals from our initial planet formation simulations (section \ref{sect:meth_form}).  Right panel: 1285 collisional fragments generated in our follow-on suite of $\sim$10$^{5}$ giant impact simulations.  Simulations considering a Neptune-mass planet at 1.0 au (``GP'') are plotted with red points.}
	\label{fig:debris}
\end{figure*}

We derive leftover planetesimal populations directly from our simulations sets that incorporate planetesimals (see table \ref{table:ics} and the left panel of figure \ref{fig:debris}), and perform an additional batch of $\sim 10^{5}$ simulations of the final few giant impacts in our systems to derive collisional fragment distributions.  This additional suite of computations utilizes a modified version of the \textit{Mercury} integration package described in \citet{chambers13} that incorporates collisional parameter space mapped in \citet{lands12,sandl12} to track hit-and-run and fragmenting impacts.  Aside from the fragmentation algorithm, the integration parameters (i.e.: time-step, ejection radius, etc.) are identical to those utilized in section \ref{sect:meth_form}.  To prevent the calculation from becoming intractable, the user must establish a minimum fragment mass \citep[$MFM$; 0.005 $M_{\oplus}$ in our simulations based off previous experience with the code:][]{quintana15,clement18_frag}.  When the algorithm detects a fragmenting collision, the total remnant mass is divided into a number of equal mass fragments with $M > M_{MFM}$ that are ejected in uniformly spaced directions in the collisional plane at $v \simeq 1.05 v_{esc}$.  If the total eroded mass is less than the $MFM$, the collision is considered totally accretionary.  While this treatment of fragmentation is obviously somewhat contrived (see section \ref{sect:discuss} for a discussion of some of these caveats), a first-order dissection of the distinctive populations of collisionally generated debris and unaltered planetesimals is important for our study as the fragment material might posses enhanced volatile and light element inventories if it is ejected from near the surface of partially or fully differentiated proto-planets.  Moreover, since we utilize these simulations to infer the orbital distribution of debris given a supposed size frequency distribution (SFD), our particular selection of MFM is not carried forward into our study of late bombardment, and is therefore only important for keeping our experiments computationally feasible.

To generate initial conditions for these simulations, we extract the state of our planet formation models at the point where $N_{emb}=$ 10 (15 for simulations that do not include planetesimals), remove all small bodies, and apply small changes ($\delta_{a} \lesssim$ 0.001) to the embryos' orbits to generate unique evolutions.  These systems are then integrated for 20 Myr.  Through this process, we generate a population of 1285 surviving collisional fragments (right panel of figure \ref{fig:debris}).

While the distributions of fragment orbital elements are relatively similar in our various simulation sets investigating different stellar and disk masses, perturbations from an external, Neptune-mass planet mildly heat the eccentricities and inclinations of the small remnant bodies compared to the simulations that do not incorporate a giant planet model.  We also find no differences between the distribution of fragments produced in systems that originated in 200 embryos compared with those beginning with 20 embryos (table \ref{table:ics}).  However, it is clear from figure \ref{fig:debris} that surviving fragments in the inner component of the disk are much rarer than surviving planetesimals.  We find this to be the result of high rates of accretion by the central body for fragments generated at low heliocentric distances \citep[by virtue of the low perihelia distances they are endowed with when ejected by our code, see also:][]{clement19_merc}.  As these fragments would likely be removed rapidly by Yarkovsky drift (see section \ref{sect:discuss}), we do not view this as a serious caveat on our results.  However, it should be noted that the exact physics underlying the process of collisional fragmentation during planetary-scale collisions is still not fully understood.  For these reasons, we elect to treat our fragment and planetesimal populations separately in the majority of our bombardment experiments.  

\subsubsection{Exo-Asteroid Belts}

The largest concentration of surviving debris material in all of our simulations lies in the vicinity of the most distant rocky planets, including a sizable number of planetesimals and fragments that are scattered into the ``exo-asteroid belt'' region exterior to our outer terrestrial disk boundary of 0.5 au \citep[e.g.:][]{izidoro16,ray17sci}.  On average, these belt-analogs possess total masses of 0.034 $M_{\oplus}$ in scattered planetesimals and 0.012 $M_{\oplus}$ in collisional fragments.  Simulations incorporating a Neptune-mass planet at 1.0 au have systematically less massive belts (an average of just $\sim$0.0064 $M_{\oplus}$ in scattered planetesimals) as a result of scattering events with the massive planet, direct accretion of planetesimals, and the presence of overlapping first order mean motion resonances near 1.0 au \citep[e.g.:][]{wisdom80,deck13}.  Conversely, integrations considering more massive central stars strand a larger fraction of material in the terrestrial and asteroid belt regions (an average of 0.076 $M_{\oplus}$ and 0.11 $M_{\oplus}$ worth of leftover planetesimals in our 0.5 $M_{\odot}$ and 0.6 $M_{\odot}$ simulations, respectively).  We attribute this result to the relatively larger orbital velocities in these simulations providing stronger gravitational scattering events that disfavor accretion.

The average total leftover fragment ($\sim$0.04 $M_{\oplus}$) and planetesimal ($\sim$ 0.06 $M_{\oplus}$) mass in our simulations (combining debris in the vicinity of the terrestrial planets with exo-asteroid belt material) is remarkably similar to that in similar simulations of terrestrial planet formation in the solar system \citep[presumably several lunar masses to provide a sufficient population of late impactors to source the late veneer, e.g.:][]{ray13,brasser16_lhb,clement18_frag}.  Systems considering an external, Neptune-mass body tend to possess lower total left-over masses in the terrestrial and exo-belt regions (for additional data, see table \ref{table:results}).  This is likely attributable to resonant perturbations from the planet exciting the velocity dispersion of the small body population, thereby providing increased high-speed collisions that disfavor perfect accretion \citep[e.g.:][]{lands12}.  However, the range of outcomes in terms of total debris mass overlap significantly in each of our simulation batches. 

\subsubsection{1 Gyr Bombardment Simulations}

We experiment with two different approaches for deriving late bombardment curves in our chosen systems of M-Dwarf hosted planets.  In both cases, we adjust the semi-major axes of all our debris particles by linearly scaling the orbital periods of our initial inner and outer disk boundaries (0.01 and 0.5 au) to the inner and outermost planet in each system. In our first experiment, we simply integrate each debris particle\footnote{Note that we combine leftover planetesimals and fragments from all of our planet formation simulations (regardless of stellar mass and giant planet inclusion) in our bombardment models because the orbital distributions of the debris generated in our different simulation suites (table \ref{table:ics}) are not particularly distinctive.} as a semi-active body in each system of interest for 1 Gyr (in practicality, we perform 52 separate simulations, each incorporating $\sim$50 different debris particles).  These simulations leverage the same $Mercury$ integrator and settings as our planet formation simulations described in section \ref{sect:meth_form}. Through this process, we generate chronological impact curves for each planet that we can calibrate by varying both the presumed initial SFD and each impactor's assumed initial composition (which we relate to its initial origin in our planet formation simulations).

To validate our above methodology, we perform a second set of similar bombardment experiments (studying our collisional fragment distributions) that incorporate much higher particle resolution by leveraging the \textit{GENGA} GPU integration package \citep[a GPU-parallelized version of the \textit{Mercury} code described in][]{genga}.  Aside from the change in integrator, the set-up for these simulations is identical to that of the models described above.  To take advantage of \textit{GENGA's} superior performance, we include 10,000 debris particles in these simulations.  The orbits of these new objects are interpolated by sampling from the original debris populations' distributions of semi-major axes, eccentricities and inclinations.  The masses of the particles are assigned in a manner to mimic the SFD of the modern asteroid belt down to 5 km objects.

\section{Results}

\subsection{Resultant Planetary Systems}

\begin{table*}
\centering
\begin{tabular}{c c c c c c c}
\hline
$M_{*}$ & $M_{Disk}$ & Neptune $@$ 1.0 au & $N_{pln}$ & $N_{HZ}$ & $M_{AB}/M_{AB,SS}$ & $AMD/AMD_{TP,SS}$  \\
\hline
0.1 $M_{\odot}$ & 3.0 $M_{\oplus}$ &  N & 6.88 & 0.88 & 66.6 & 2.28\\
 &  &  Y & 6.0 & 0.77 & 1.66 & 7.29\\
 & 6.0 $M_{\oplus}$ &  N & 6.77 & 1.0 & 40.0 & 4.39\\
 &  &  Y & 5.55 & 1.0 & 0.0 & 6.09\\
0.2 $M_{\odot}$ & 3.0 $M_{\oplus}$ &  N & 8.88 & 1.11 & 81.6 & 1.90\\
 &  &  Y & 8.77 & 0.77 & 3.33 & 1.87\\
 & 6.0 $M_{\oplus}$ &  N & 7.55 & 1.44 & 46.6 & 2.70\\
 &  &  Y & 7.11 & 1.22 & 0.0 & 3.26\\
0.3 $M_{\odot}$ & 3.0 $M_{\oplus}$ &  N & 10.2 & 1.22 & 70.0 & 1.19\\
 &  &  Y & 9.66 & 1.22 & 3.33 & 2.06\\
 & 6.0 $M_{\oplus}$ &  N & 7.88 & 1.66 & 73.3 & 1.72\\
 &  &  Y & 8.88 & 1.77 & 0.0 & 1.87\\
0.4 $M_{\odot}$ & 3.0 $M_{\oplus}$ &  N & 10.5 & 1.22 & 36.6 & 0.95\\
 &  &  Y & 10.6 & 1.33 & 10.0 & 0.83\\
 & 6.0 $M_{\oplus}$ &  N & 9.55 & 1.77 & 66.6 & 1.28\\
 &  &  Y & 9.55 & 1.66 & 3.33 & 1.44\\
0.5 $M_{\odot}$ & 3.0 $M_{\oplus}$ &  N & 11.5 & 1.22 & 36.6 & 0.82\\
 &  &  Y & 10.6 & 1.55 & 3.33 & 0.63\\
 & 6.0 $M_{\oplus}$ &  N & 10.6 & 2.12 & 56.2 & 0.96\\
 &  &  Y & 9.77 & 1.55 & 6.66 & 1.75\\
0.6 $M_{\odot}$ & 3.0 $M_{\oplus}$ &  N & 12.8 & 1.0 & 33.7 & 0.43\\
 &  &  Y & 11.8 & 0.66 & 13.3 & 0.54\\
 & 6.0 $M_{\oplus}$ &  N & 10.3 & 1.88 & 73.3 & 0.98\\
 &  &  Y & 10.6 & 2.33 & 6.66 & 0.65\\
\hline
\end{tabular}
\caption{Summary of final system properties in our various rocky planet formation simulations (table \ref{table:ics}).  The columns are as follows: (1) The mass of the central star in solar units, (2) the initial mass of the terrestrial-forming disk in Earth units, (3) whether or not the system contains a Neptune-mass giant planet at 1.0 au, (4) the mean number of planets formed per system, (5) the mean number of planets with $M>$ 0.5 $M_{\oplus}$ formed in the HZ per system, (6) the average total mass of leftover planetesimals and/or embryos in the exo-asteroid belt region ($a>$0.5 au) normalized to the total mass of the modern asteroid belt, and (6) the final system AMD (equation \ref{eqn:amd}) normalized to the value for Mercury, Venus, Earth and Mars.}
\label{table:results}
\end{table*}

Table \ref{table:results} summarizes several properties of our fully accreted systems of rocky planets.  In general our simulations indicate that, provided M-Dwarf-hosted short-period planets form in-situ from relatively massive disks, HZ planets are essentially ubiquitous.  Indeed, regardless of the parameters varied, our simulations systematically yield $\sim$1-2 planets with $M>$ 0.5 $M_{\oplus}$ in the conservative HZ \citep[note, however, that $M \lesssim$ 0.01 $M_{\oplus}$ water worlds might also be habitable:][]{goldblatt15,arnscheidt19}.  However, we remind the reader that our results might be biased by the low resolution of our numerical models.  While our simulations considering less massive, $M_{Disk}=$ 3.0 $M_{\oplus}$ populations of embryos and planetesimals struggle to form $\sim$Earth-mass planets in the HZ by virtue of their lower initial surface density profiles, they also form more total planets as the lower individual masses render more compact dynamical configurations stable \citep{chambers96}.  

\begin{figure}
	\includegraphics[width=.5\textwidth]{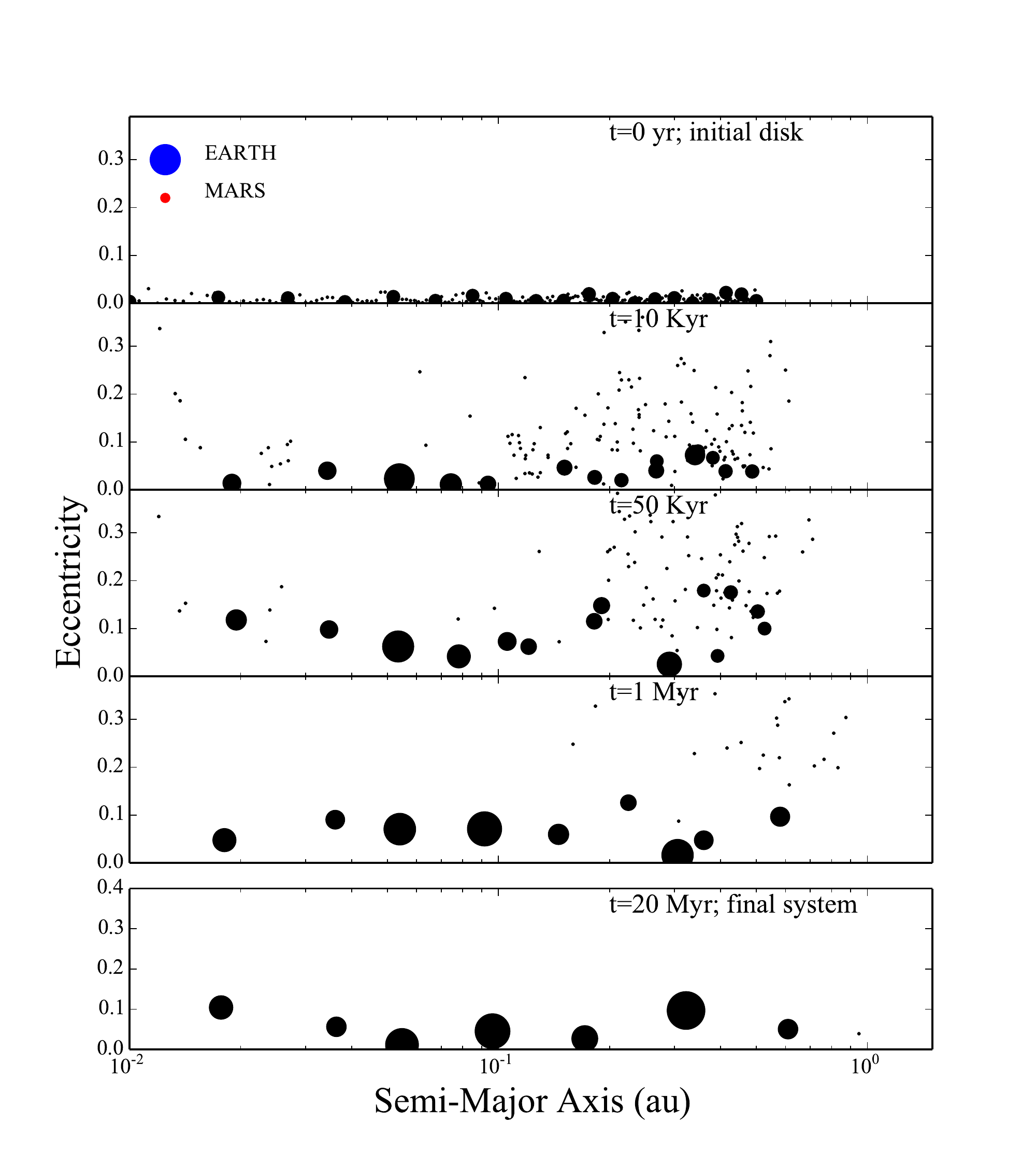}
	\caption{Example evolution of a system in our $M_{*}=$ 0.1 $M_{\odot}$, $M_{disk}=$ 6.0 $M_{\oplus}$, no giant planet batch of simulations that produced a reasonable analog of the TRAPPIST-1 system \citep{gillon16}.  The semi-major axis and eccentricity of each simulation body are plotted, and the size of each point is proportional to the particle's mass (comparative points for an Earth and Mars-mass are provided in the top panel.  The final planets' masses are 0.53, 0.36, 0.65, 1.1, 1.2, 1.4 and 0.36 $M_{\oplus}$, respectively.  For comparison, the mass of the TRAPPIST-1 planets with increasing semi-major axis are: 1.374$\pm$0.069, 1.308$\pm$0.056, 0.388$\pm$0.012, 0.692$\pm$0.022, 1.039$\pm$0.031, 1.321$\pm$0.038 and $\pm$0.326$\pm$0.020 $M_{\oplus}$ \citep{trappist_new}.}
	\label{fig:trappist}
\end{figure}

Unsurprisingly, forming systems of multiple Earth-mass planets in the HZ like those around TRAPPIST-1 \citep[or even Super-Earths like HIP 38594b:][]{feng20b} in-situ axiomatically requires a more massive disk of embryos planetesimals.  While future observations characterizing the short-period populations of $\sim$0.01-0.5 $M_{\oplus}$ planets will be important for constraining the relative importance of planetesimal accretion and giant impacts in the formation HZ planets around low-mass stars, our simulations demonstrate it to be a plausible genesis scenario.  Indeed, several of our evolved systems reasonably replicate the architectures of detected systems of real exoplanets.  Figure \ref{fig:trappist} depicts an example evolution of such a system from our $M_{*}=$ 0.1 $M_{\odot}$, $M_{disk}=$ 6.0 $M_{\oplus}$, no giant planet batch that yields a system of seven $\sim$0.4-1.4 $M_{\oplus}$ bodies similar to the TRAPPIST-1 planets.  While the radial spacing of these planets is less compact than that of TRAPPIST-1 \citep[as those worlds are suspected to be locked in an 8:5;5:3;3:2;3:2;4:3;3:2 resonant chain:][]{luger17_res}, we will show in the subsequent section that the dynamical offsets between neighboring planets in our systems are consistent with M-Dwarf hosted exoplanet demographics.  Moreover, it is likely that our gas-free simulations do not adequately capture the process of in-situ formation.  We plan to explore this with a forthcoming study incorporating analytical treatments for Type-I migration \citep{tanaka04} and aerodynamic drag \citep{adachi78}

Our results (table \ref{table:results}) elucidate a clear trend of simulations incorporating a Neptune-mass planet at 1.0 au finishing with less massive inventories of small bodies that could potentially source sufficiently prolonged volatile delivery to desiccated HZ planets.  As we discuss in section \ref{sect:meth_bomb}, this is largely a result of scattering events and direct accretion by the giant planet.  When we include an additional planet, the resultant exo-asteroid belts have masses comparable to that of the modern solar system \citep[note that this is consistent with the so-called empty primordial asteroid belt hypothesis for the solar system:][]{izidoro15,izidoro16,ray17sci}.  Conversely the typical leftover asteroidal mass in simulations without giant planets exceed that of the solar system by around two orders of magnitude.  This demonstrates an important result, as the presence of powerful resonances with external giant planets would by required to consistently deliver material to HZ planets over Gyr timescales \citep[e.g.:][]{milani90,bottke15,dencs19}.  Therefore, at first glance we expect that HZ planets formed in-situ in systems harboring long-period giant planets might experience delivery rates of volatile-rich small bodies similar to the early bombardment of the Earth (albeit over a different dynamical timescale by virtue of the shorter orbital periods).  Conversely, the systems more likely to host massive inventories of water-ice-rich objects capable of sourcing delayed volatile delivery might lack the massive planets needed to perturb the material onto HZ-crossing orbits.

\subsubsection{Dynamical configurations}

\begin{figure}
	\centering
	\includegraphics[width=.5\textwidth]{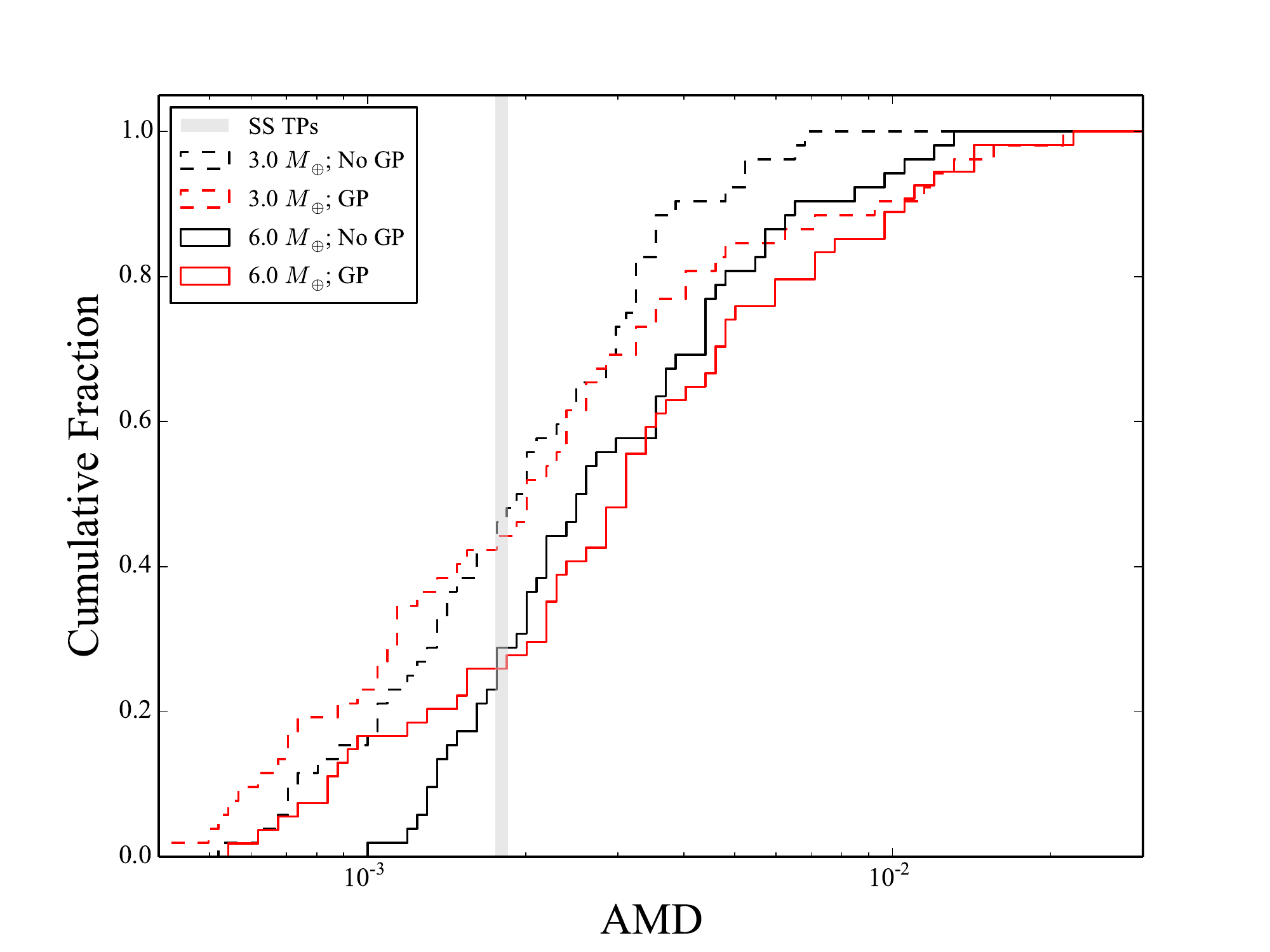}
	\caption{Cumulative distribution of system angular momentum deficits in our various planet formation simulations investigating different disk masses (3.0 and 6.0 $M_{\oplus}$; dashed and solid lines, respectively) and giant planet models (No giant planets and a Neptune-mass planet at 1.0 au: black and red lines, respectively).  The grey vertical line denotes the solar system value for Mercury, Venus, Earth and Mars.}
	\label{fig:amd}
\end{figure}

Dynamically speaking, our numerically generated systems are qualitatively similar to known multi-planet transiting systems, and the solar system's terrestrial worlds.  We quantify the degree of dynamical excitation in our systems by calculating the angular momentum deficit (AMD) of all planets with $M>$ 0.05 $M_{\oplus}$ \citep{laskar97}:

\begin{equation}
	AMD = \frac{\sum_{i}m_{i}\sqrt{a_{i}}[1 - \sqrt{(1 - e_{i}^2)}\cos{i_{i}}]} {\sum_{i}m_{i}\sqrt{a_{i}}} 
	\label{eqn:amd}
\end{equation} The AMD of a system essentially gauges the degree to which the collection of orbits deviate from that of a perfectly circular, co-planar system.  Figure \ref{fig:amd} plots the cumulative distribution of AMDs in our various simulation suites testing different values of $M_{Disk}$ and giant planet models.  While the inclusion of an external massive planet (on a circular orbit) has a negligible effect on the resulting planetary system's dynamical excitation, the difference in AMDs between our simulation sets testing $M_{Disk}=$ 3.0 and 6.0 $M_{\oplus}$ demonstrates the tradeoffs of forming HZ planets in-situ from a massive disk.  While a high initial surface density of solids is essential for growing planets of order the mass of the Earth, such concentrated regions of proto-planets naturally produce systems of fully evolved planets on more eccentric and inclined orbits.  Thus, the HZ planets themselves would likely undergo larger magnitude secular oscillations \citep[most consequentially in obliquity and perihelia, e.g.:][]{berger78,laskar93} that change their climates over kyr-Myr timescales, and would also be more susceptible late planetary collisions and ejections \citep[e.g.:][]{laskar09,batygin15}.  While these phenomena might present challenges for the emergence of life, it is important to note that models of terrestrial planet formation in the solar system incorporating numerical methodologies akin to that presented here systematically struggle to replicate the low eccentricities of Earth and Venus \citep[e.g.:][]{chambers98,ray06,clement21_emb}.  Moreover, it is still unclear whether short-period rocky planets around M-Dwarfs universally possess low eccentricities and inclinations \citep[as expected for a resonant system like TRAPPIST-1:][]{grimm18}, or more moderate values capable of driving large obliquity and perihelia cycles \citep[e.g.:][]{shan18,quarles20}.

\begin{figure}
	\centering
	\includegraphics[width=.5\textwidth]{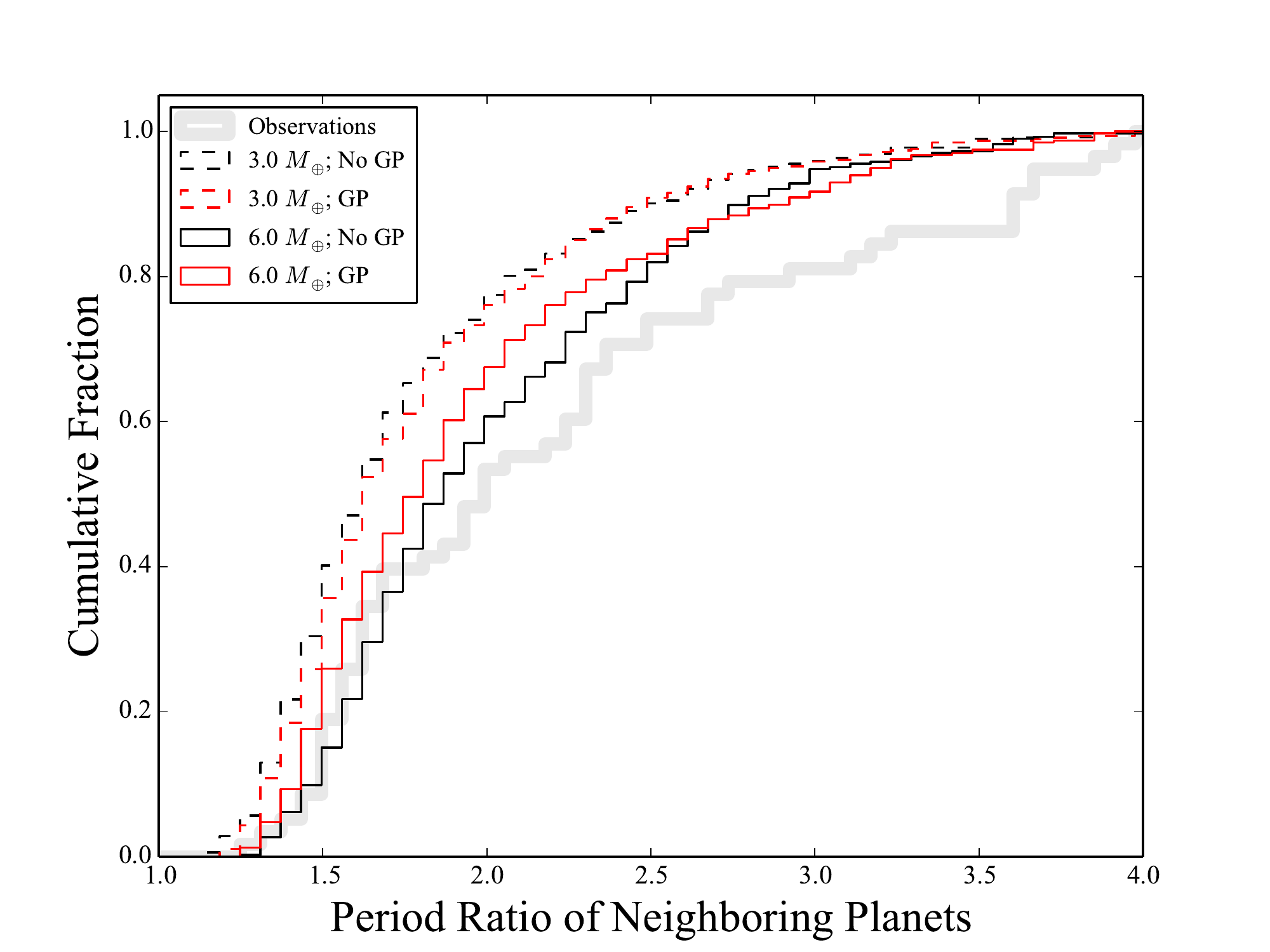}
	\caption{Cumulative distribution of neighboring planet ($M>$ 0.10 $M_{\oplus}$) orbital period ratios in our various planet formation simulations investigating different disk masses (3.0 and 6.0 $M_{\oplus}$; dashed and solid lines, respectively) and giant planet models (No giant planets and a Neptune-mass planet at 1.0 au: black and red lines, respectively).  The grey line plots the period ratio of all multi-planet systems of M-Dwarf-hosted ($M_{*}<$ 0.6 $M_{\odot}$) planets with $a<$ 0.5 au in NASA's exoplanet archive (queried on 2 July 2021 for the preparation of this manuscript).}
	\label{fig:perrat}
\end{figure}

Figure \ref{fig:perrat} plots the cumulative distribution of neighboring planet period ratios in our various simulation sets against those of all M-Dwarf-hosted multi-planet systems.  Here, we only consider exoplanets with $a<$ 0.5 au and $R<$ 2.0 $R_{\oplus}$ to more accurately compare with our modeled scenario.  It is clear that our simulations considering disk masses of 6.0 $M_{\oplus}$ provide a reasonable match to the observed period ratios of exoplanets, however it is important to note that these observations are biased due to potentially undetected planets.  In particular, undetected small planets \citep[e.g.:][]{faridani21} similar to those formed in our $M_{Disk}=$ 3.0 $M_{\oplus}$ simulations would likely shift the observed curve left by virtue of potentially inhabiting more compact orbital architectures.  Nevertheless, figure \ref{fig:perrat} depicts a reasonable match to the observed pile-up of observed systems with orbital period ratios near, but not exactly within the 3:2 MMR.  This perceived under-abundance of resonant planets compared to predictions from convergent migration models \citep[namely applied to Super-Earths orbiting Sun-like stars, e.g.:][]{ogihara15,ogihara18} has also been invoked to suggest that such planets acquired their orbital architectures via delayed episodes of dynamical instability \citep{izidoro17,izidoro19} after forming in quasi-stable resonant configurations \citep[e.g.][]{volk15}. While such a formation model is certainly plausible for explaining chains of resonant and non-resonant short-period planets around M-Dwarfs like TRAPPIST-1 and Kepler-186, our simulations indicate that in-situ formation is also capable of reconciling the period ratios of observed planets.

\subsubsection{Volatile enrichment of the HZ}
\label{sect:water}

\begin{figure}
	\centering
	\includegraphics[width=.5\textwidth]{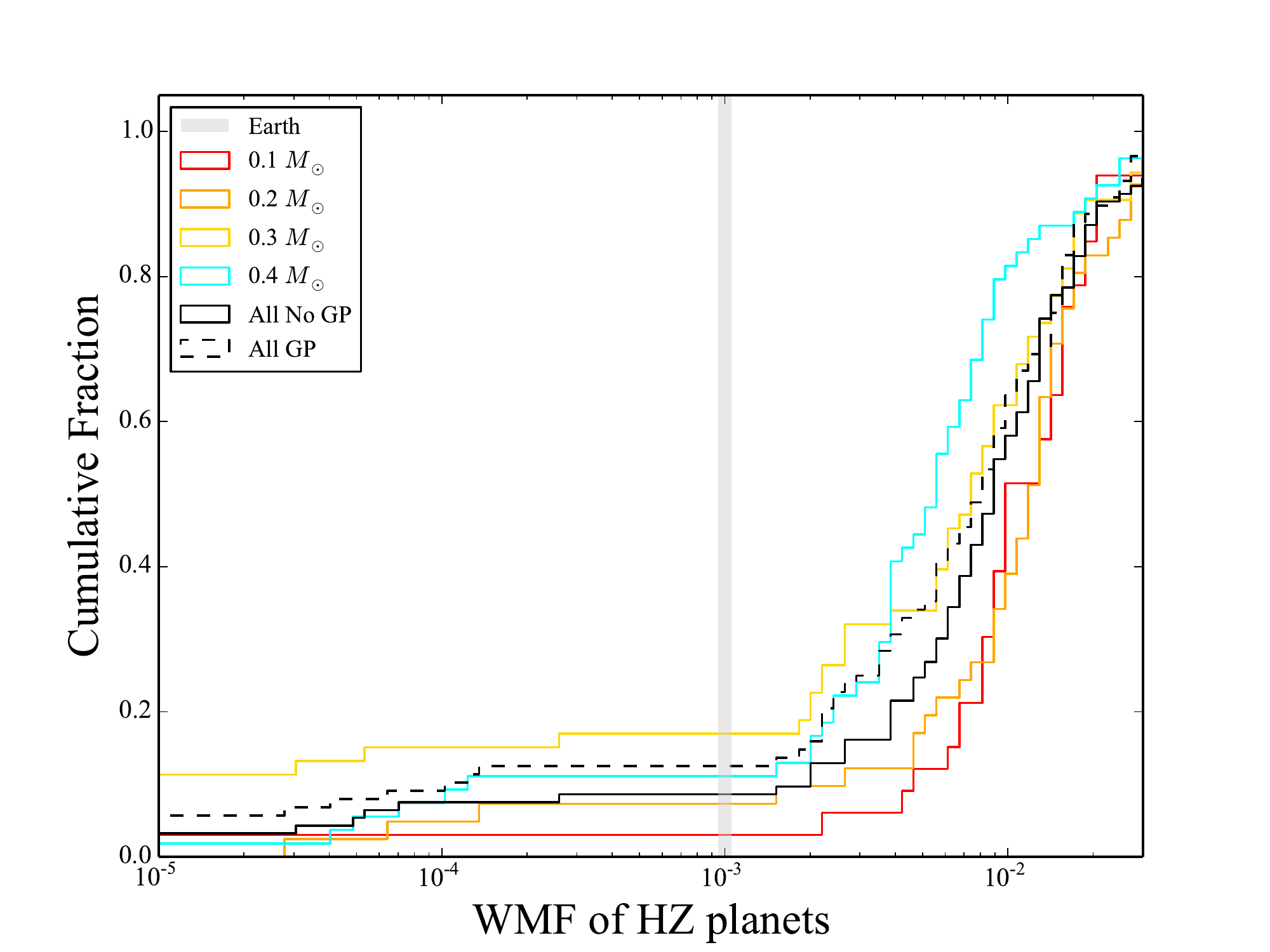}
	\caption{Cumulative distribution of final water mass fractions of HZ planets \citep[defined here by the conservative limits established in][]{hz} formed in our simulations investigating different stellar masses (different line colors).  Additionally, the dashed and solid black lines combine data for all simulations with and without a Neptune-mass planet at 1.0 au, respectively.  The grey vertical line denotes Earth's presumed bulk water content \citep[e.g.:][]{meech19}.}
	\label{fig:wmf}
\end{figure}

Our simulated terrestrial disks experience moderate to strong radial mixing of material that systematically boosts the water-ice contents of HZ planets.  Specifically, $\gtrsim$83$\%$ of HZ planets in each of our simulation sets where the snowline is located interior to our outer disk boundary accrete at least one object from beyond the line.  To quantify this enhancement, we crudely assume that all planetesimals and embryos originating inside of 0.8$a_{snow}$ possess initial WMFs of 10$^{-5}$, those between 0.8 and 1.0 $a_{snow}$ are endowed with 0.1$\%$ of their mass in the form of water, and exterior disk objects begin with 10$\%$ of their mass in the form of water-ice.  We discuss the limitations of these assumptions in section \ref{sect:meth_form}.  Figure \ref{fig:wmf} plots the cumulative distribution of final WMF for all HZ planets formed in our our $M_{*}=$ 0.1, 0.2, 0.3 and 0.4 $M_{\odot}$ simulation batches (note that the snowline is beyond the edge of our initial terrestrial disk in our investigations of more massive M-Dwarfs).  The boosted WMFs in simulations investigating smaller stars thus demonstrates the ability of HZ planets to accrete material from the more distant regions of their initial planet forming disks (i.e.: WMFs are not boosted simply as a result of accretion from the edge of the snowline).  Indeed, we note multiple incidences of HZ planets in our 0.1 $M_{\odot}$ batch with 0.025 $<a<$ 0.057 au accreting planetesimals initialized at semi-major axes beyond 0.3 au.  Overall, the radial mixing in our simulations is similar to that noted in classic models of terrestrial planet formation in the solar system employing similar disk setups to our models \citep[e.g.:][]{ray04,ray07,obrien18}.

The large pileup of Earth analogs with water contents of $\sim$1$\%$ corresponds to objects that accrete exactly one embryo from beyond the snow line.  Assuming a nominal $\sim$1.0 $M_{\oplus}$ planet in our batch of 6.0 $M_{\oplus}$ disks (or, equivalently, a $\sim$0.5 $M_{\oplus}$ planet in our set of 3.0 $M_{\oplus}$ disks), an impact from a water-ice rich embryo would deliver approximately 14$\%$ of the Earth analogs' ultimate mass.  If the impactor possessed a water-ice content of 10$\%$, the resulting HZ planet's WMF would be approximately 10$^{-2}$.  Thus, the final water contents themselves are essentially an artifact of our initial disk's presumed compositional structure, and the more important result is the observed strong radial mixing.

While our numerical estimates of each HZ planets' final WMF are obviously contrived, it is clear that Earth-analogs formed in-situ around M-Dwarfs might be endowed with initially high volatile inventories when compared to the modern Earth and Mars.  Thus, it might be possible for such planets to replenish atmospheres desiccated during their host stars lengthy cooling epochs via volcanic outgassing of water-rich mantle material \citep[e.g.:][]{moore20}.  While the presence of an external giant planet slightly curtails this incorporation of water-rich material, the effect is fairly minor.  Moreover, it is important to note that the more dynamically compact configurations significantly bolster the chances of a HZ planet accreting material from beyond the snowline.   While a planetesimal in the solar nebula located at 2.8 au must attain $e\simeq$ 0.66 \citep[an unlikely occurrence in models of terrestrial planet formation:][]{clement19_merc} to be directly accreted by the young Earth without undergoing a series of fortuitous scattering events, a similarly situated icy planetesimal around an 0.1 $M_{\oplus}$ M-Dwarf at 0.085 au could be accreted by a HZ planet at 0.06 au if it achieves an eccentricity of around 0.3.  While this analysis is clearly complicated by the relative location of the snowline during the enhanced luminosity epoch of these star's early lives, this might not pose a significant issue if planetesimals and embryos form further out in the disk before migrating inwards.

\subsection{Late Bombardment}
\label{sect:bomb}

\subsubsection{Rapid Debris Removal}

\begin{figure*}
	\centering
	\includegraphics[width=.49\textwidth]{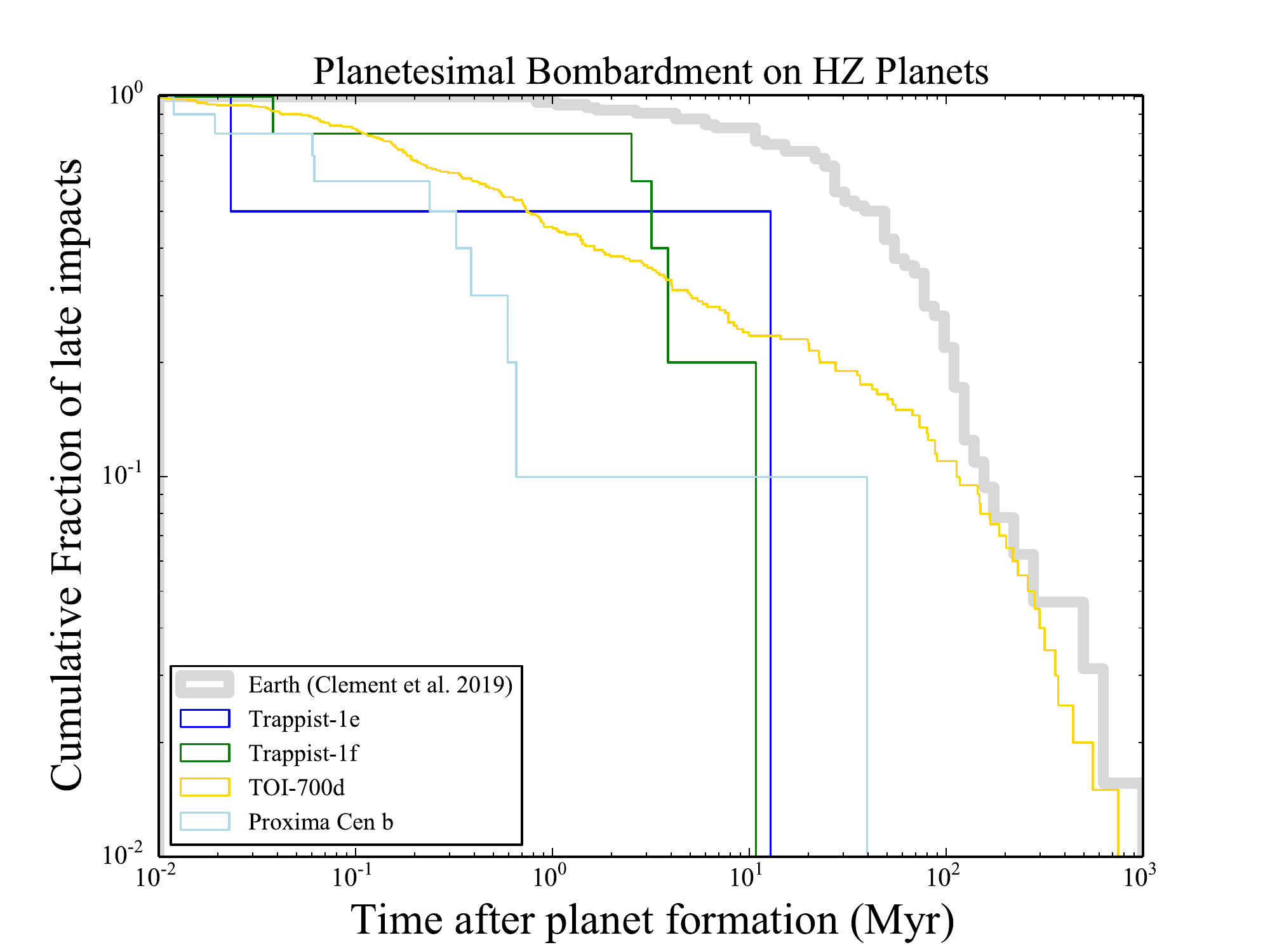}
	\includegraphics[width=.49\textwidth]{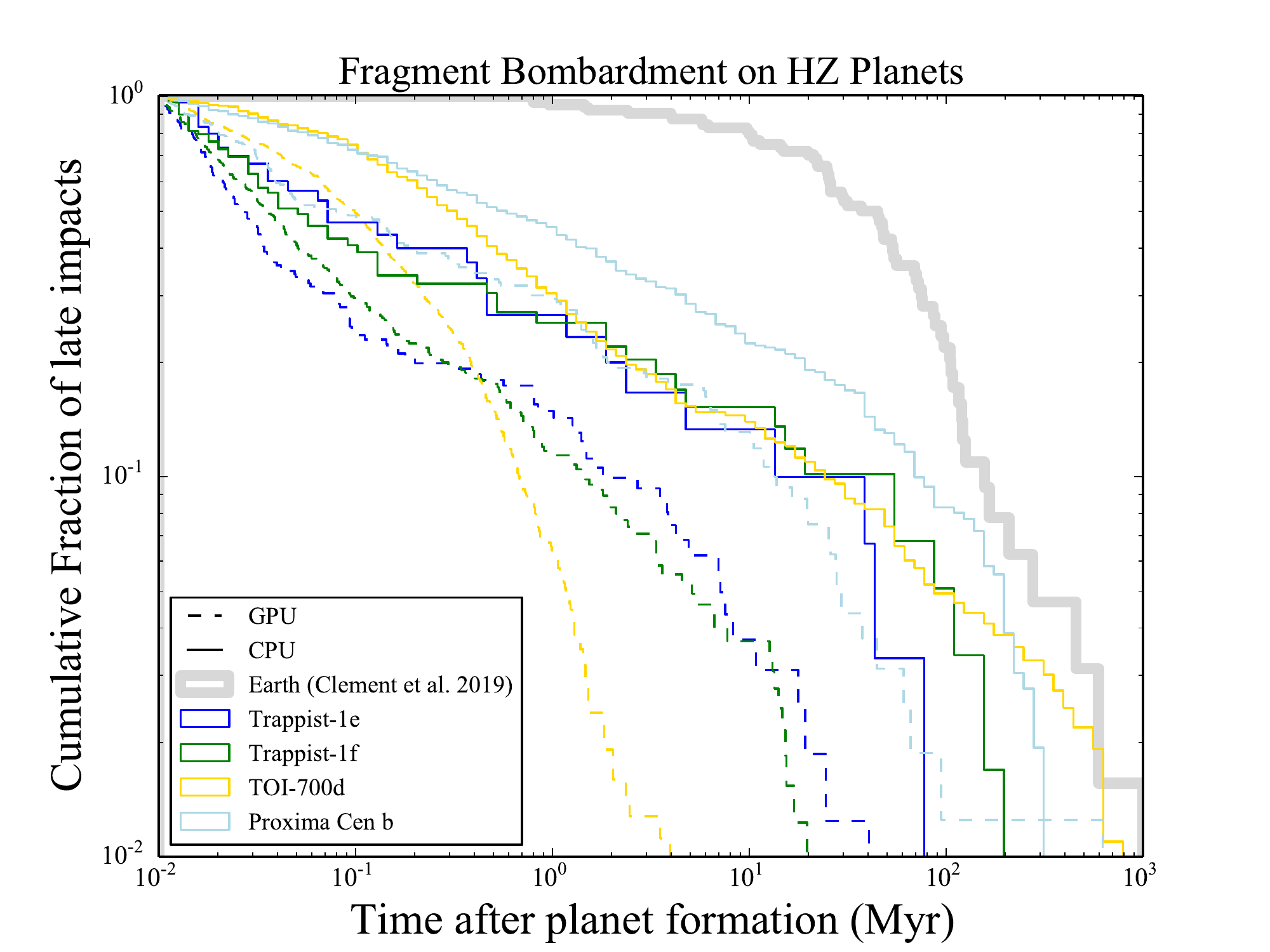}
	\caption{Temporal distribution of material delivery in our various bombardment simulations investigating remnant planetesimal (left panel) and collisional fragment (right panel) bombardment of HZ planets.  Time is plotted with respect to the end of the planet formation process (20 Myr for our M-Dwarf models and 200 Myr for our solar system comparison case).  Each histogram is weighted by the total number of planetesimals and fragments simulated in each system.  The grey line plots similar simulations investigating late bombardment on Earth in \citet{clement18_frag}.  Additionally, the right panel compares our \textit{Mercury}-derived bombardment curves with those computed in a single \textit{GENGA}-simulation.}
	\label{fig:impacts_raw}
\end{figure*}

We develop a semi-analytic model to analyze whether our extended simulations of debris evolution in TRAPPIST-1, Proxima Centauri and TOI-700 support the occurrence of late, large, wet impacts on the star's respective HZ planets.  The raw data (cumulative fraction of total system debris accreted by each planet of interest) is plotted for our \textit{Mercury} CPU and \textit{GENGA} GPU simulations in figure \ref{fig:impacts_raw}.  For the purposes of our subsequent analysis, we focus on the TRAPPIST-1 planets in the conservative HZ \citep[e and f:][]{hz}, though we find that TRAPPIST-1g (situated in the optimistic HZ) accretes a similar number of remnant particles at a similar rate.  We also compare our measured bombardment rates to those for the young Earth extrapolated from a similar study of planet formation in the solar system presented in \citet{clement18_frag}.  Notably, the computations in that paper utilized the same code and fragmentation algorithm employed here.  Specifically, we integrate each remnant planetesimal and collisional fragment from all "instability" simulations that finish with the Jupiter-Saturn period ratio less than 2.8 in the presence of the modern solar system \citep[these model the formation of the terrestrial planets occurring in conjunction with the so-called Nice Model instability, and have been demonstrated success when measured against a number of observational constraints:][]{Tsi05,clement18,deienno18,mojzsis19,brasser20,nesvorny21_tp}.

The right panel of figure \ref{fig:impacts_raw} compares the results of our high-resolution GPU simulations (dashed lines) with our co-added CPU simulations.  While the agreement between the respective curves for each planet is reasonable, it is clear that our GPU simulations tend to produce fewer late impacts than our CPU runs (the largest difference can be seen in the yellow curves for TOI-700).  We attribute this result to the more realistic size distribution assumed in our GPU simulations (in our CPU simulations we assign all debris particles equal masses), however we do not test this hypothesis with additional simulations.  Nevertheless, it is important to note that, by focusing on the bombardment curves derived from our co-added CPU simulations in our subsequent analyses, we are essentially assuming a best-case-scenario in order to place upper limits on the late bombardment flux in our systems.

\begin{figure}
	\centering
	\includegraphics[width=.5\textwidth]{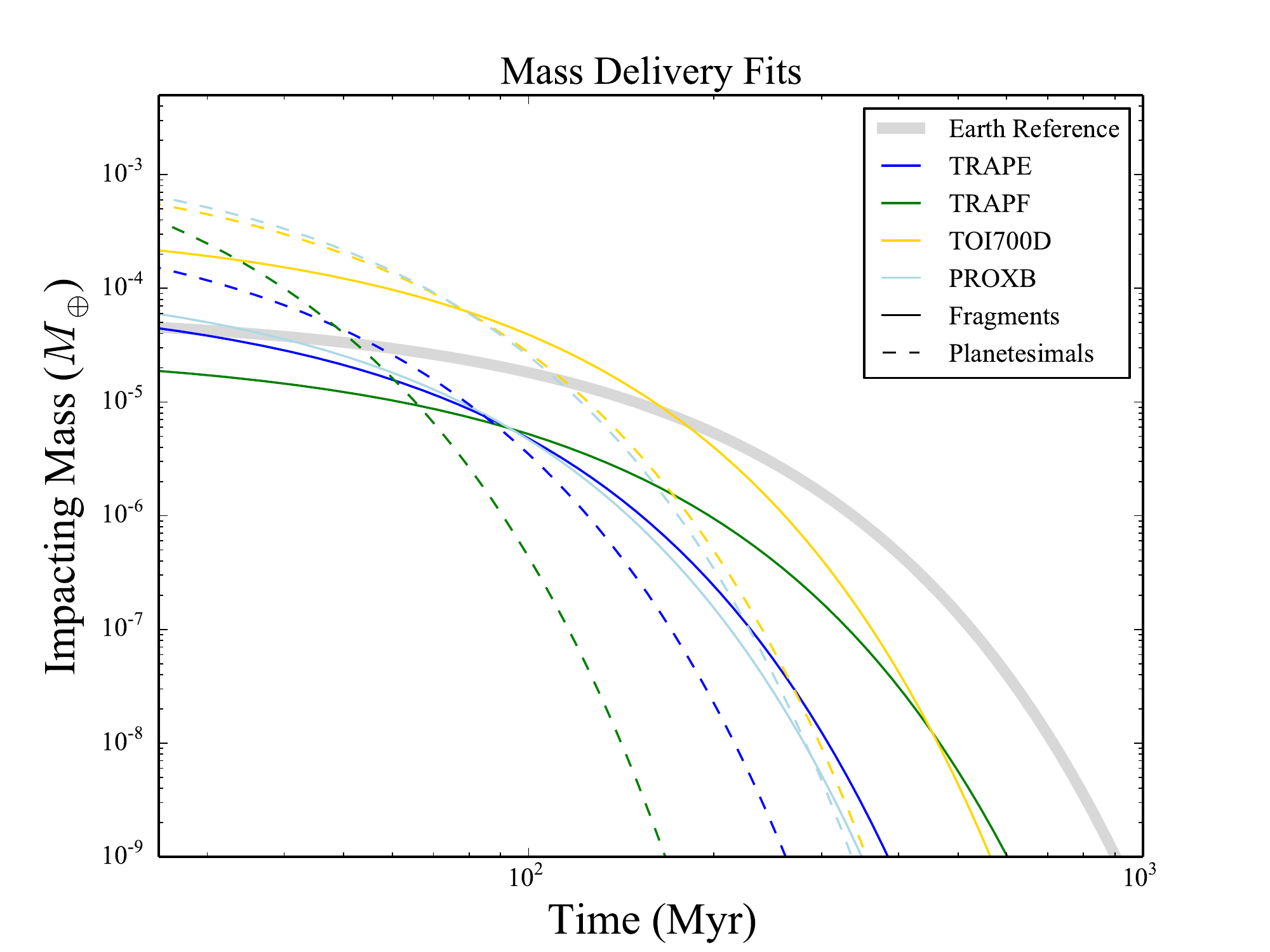}
	\caption{Exponential decay fits of the bombardment chronologies (probability distribution function) plotted in figure \ref{fig:impacts_raw}.  Time is plotted with respect to the end of the planet formation process (20 Myr for our M-Dwarf models and 200 Myr for our solar system comparison case).  Here, we weight each curve by the presumed total leftover mass (derived as the average leftover mass in the respective source planet formation simulations).  Specifically, we assume that our reference model for the Earth \citep{clement18_frag} accretes material from a debris populations with a total mass of 0.10 $M_{\oplus}$, and our various M-Dwarf-hosted debris field contain 0.04 $M_{\oplus}$ in collisional fragments and 0.06 $M_{\oplus}$ in planetesimals.  Thus, differences in the areas under these curves reflect differences in the total mass accreted by each planet after the conclusion of the planet formation process.}
	\label{fig:impacts_curves}
\end{figure}

From our raw impact chronologies (figure \ref{fig:impacts_raw}), we fit exponential decay curves and utilize their respective probability distribution functions as inputs for our analytic experiments.  Examples of these curves are plotted in figure \ref{fig:impacts_curves}.  We also generate a series of similar curves for eight separate radial origin bins to determine the probability of accreting an object from a particular initial location in the disk at any given time.  As expected given the shorter orbital periods of HZ planets around low-mass stars, debris is removed significantly faster in our systems of interest than in the solar system model.  In this manner, the planets around TOI-700 (a 0.41 $M_{\odot}$ star) experience delayed episodes of bombardment more similar to that of the young Earth than those of our other systems' HZ planets.  We also find that our planetesimal populations are removed far quicker than the collisional fragments.  We attribute this dissimilarity to the differences in orbital distributions of the debris plotted in figure \ref{fig:debris}.  Our fragment distributions span a radial range that extends from the exo-asteroid belt region inwards towards the outer edge of the HZ.  Conversely, our remnant planetesimal population possess two components: (1) un-accreted objects that survive the giant impact phase on cold orbits in between planets at small semi-major axes and (2) implanted asteroids on relatively stable orbits with large semi-major axes outside of the planetary regime.  While the first population is accreted rather easily in the first few tens of kyr of our bombardment simulations, the second population is dynamically detached from the orbits of TOI-700d, TRAPPIST-1h and Proxima Centauri b Thus, in absence of perturbations from distant giant planets and non-Keplerian accelerations \citep[such as Yarkovsky drift, see discussion in section \ref{sect:discuss} and:][]{vokrouhlicky98,bottke01,dencs19,ray21}, the HZ planets in our planetesimal bombardment simulations are unable to access the majority of the asteroid belt material.  While the Proxima Centauri system does possess a Super-Earth at 1.5 au (outside of this hypothetical asteroidal reservoir), we find it to be too distant and too small to perturb asteroids onto orbits that intersect the HZ.

With the rate of material delivery from each initial radial region of our system's presumed primordial terrestrial forming disk defined, we need only input a SFD for the debris field in order to generate hypothetical bombardment chronologies.  While this analysis assumes that these planets formed in a manner similar to that modeled in our terrestrial formation simulations (section \ref{sect:meth_bomb}), we argue in more detail in section \ref{sect:discuss} that the generation and stranding of debris should occur in these systems regardless of where the planets form in the disk.  Moreover, since our simulations that do not include planetesimals generate collisional fragments at equal rates to our integrations incorporating a sea of such small bodies, it seems reasonable that Earth-size planets implanted in the HZ from the more distant regions of the disk would still be surround by debris fields if they experience a few giant impacts with other proto-planets en-route to achieving their final orbital architectures.

While the precise SFD of remnant objects is difficult to constrain in the solar system \citep[as it is tied to the size-dependent efficiency of planetesimal formation, e.g.:][]{morby09,johansen15}, the SFD of the modern asteroid belt serves as a useful proxy for our current purposes \citep[e.g.:][]{sinclair20}.  However, the SFD of asteroids that do not belong to collisional families \citep{delbo17,delbo19} seems to suggest that $D\sim$100 km objects were far more prevalent in the primordial belt than in the modern one (thus potentially evidencing the efficient formation of larger, $D\sim$100-1,000 km planetesimals).  Therefore, we also experimented with both debris populations composed entirely 100 km objects, and one that combines the main belt's SFD with an additional 100 km component and found that the different presumed distributions only slightly altered our results.

\begin{figure*}
	\centering
	\includegraphics[width=.75\textwidth]{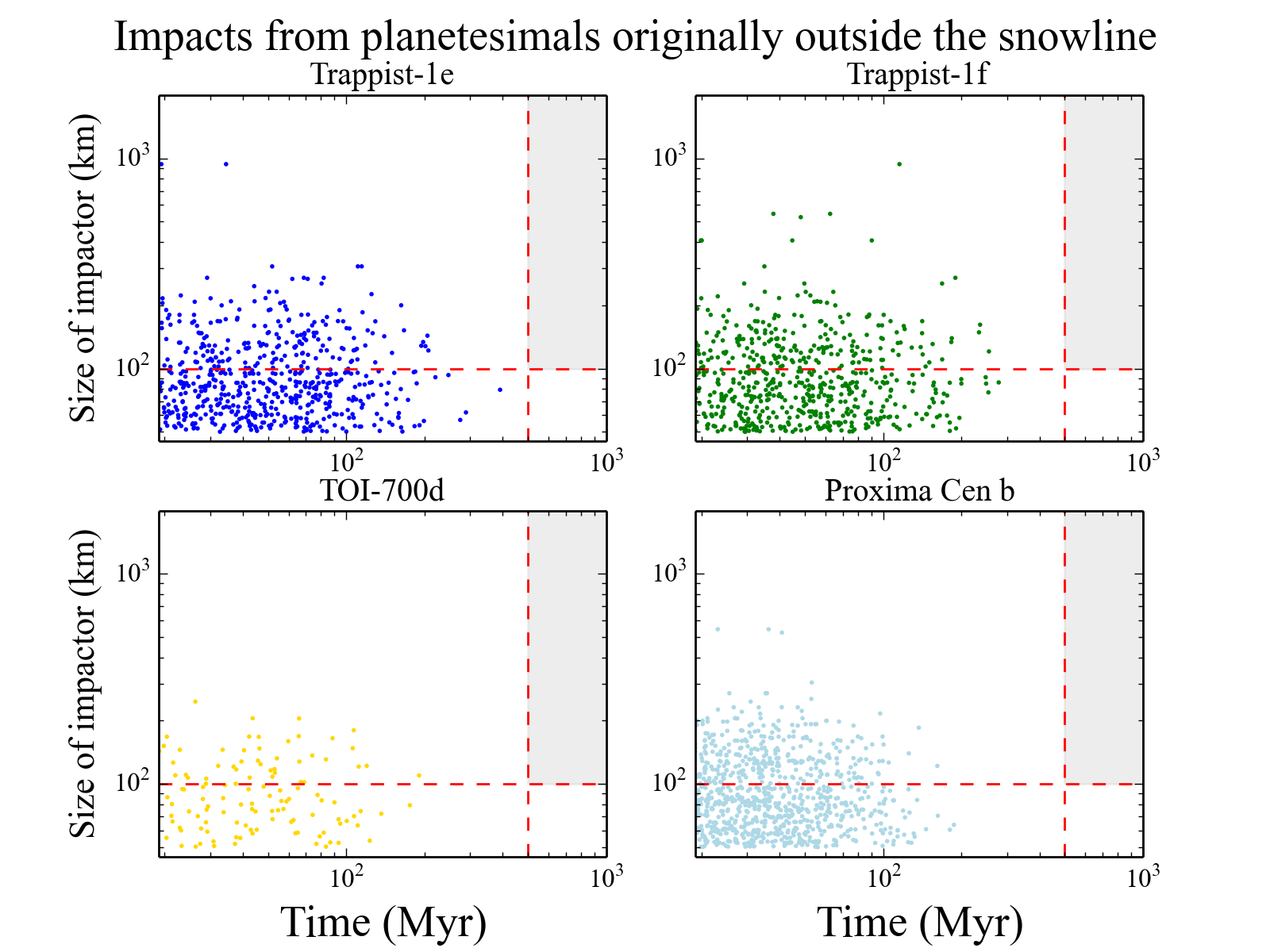}
	\caption{Derived fragment impact chronologies (here we plot all impacts of $D>$ 50 km objects) on the M-Dwarf-hosted, HZ planets TRAPPIST-1e, TRAPPIST-1f, TOI-700d and Proxima Centauri b.  All impactors (regardless of origin) are plotted.  Time is plotted with respect to the end of the planet formation process (20 Myr).  The red dashed lines and grey shaded regions represent our determined limits for achieving an impact potentially capable of reconstituting an atmosphere on a desiccated planet: $D\geq$ 100 km and $t\geq$ 500 Myr.}
	\label{fig:impacts_plns}
\end{figure*}

Figure \ref{fig:impacts_plns} plots hypothetical remnant planetesimal impact chronologies for our four HZ, Earth-size planets of interest.  For the scenario of secondary atmospheric development via small body delivery to be successful, we require our models produce at least one impact of an object with $D\geq$ 100 km that originated outside the snowline at $t\geq$ 500 Myr.  Our size requirement is based off an impactor of roughly carbonaceous chondritic composition (and a WMF of $\sim$0.10 as discussed in section \ref{sect:meth_form}) possessing roughly one Earth atmosphere worth of volatiles ($\sim 5\text{x}10^{20}$ kg).  Our time limit is roughly related to the time after star formation that the outer region of an 0.1 $M_{\odot}$ star's main sequence HZ is habitable \citep[e.g.:][]{bolmont17}.  While our reference solar system models from \citet{clement18_frag} typically yield several $D\gtrsim$ 100 km impactors that originated outside of the snowline at $t\gtrsim$ 500 Myr, such events are non-existent in our M-Dwarf simulations \footnote{Note that this flux in the solar system is consistent with chronologies derived empirically from the crater record \citep[e.g.:][]{minton19}.  The Imbrium basin, for instance, is thought to have been formed around 450 Myr after planet formation from a collision with a $D\sim$ 100 km object \citep{schultz16}.}.  Indeed, the tail-end phase of bombardment is mostly complete in each of our systems of M-Dwarf-hosted planets around $t\simeq$ 100 Myr.  While collisions persist over $\sim$Gyr timescales in our systems, our model finds that such events are statistically far more likely to consist of small projectiles that originated inside of the snowline.  We also looked at the possibility that Lidov-Kozai \citep{lidov,kozai} cycles of high-inclination exo-belt objects might lengthen dynamical lifetimes and prolong bombardment, however the characteristic timescale for such oscillations in our investigated systems is quite short:
\begin{equation}
	\tau_{Kozai} \simeq  \frac{P_{2}^{2}}{P_{1}} (1- e_{2}^{2})^{3/2}
\end{equation} 

\subsubsection{Prolonged Fragment Bombardment}

\begin{figure*}
	\centering
	\includegraphics[width=.75\textwidth]{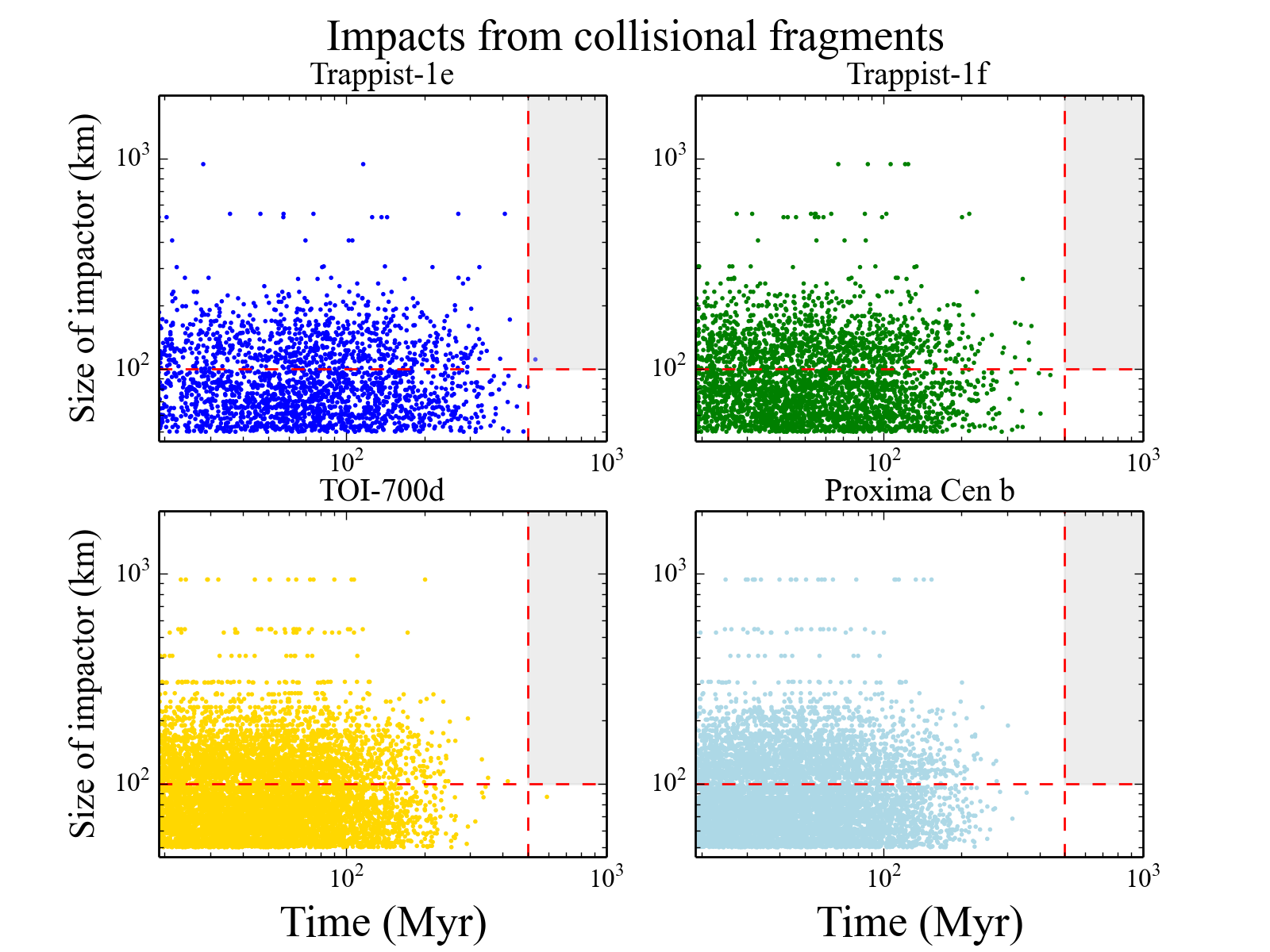}
	\caption{Derived planetesimal impact chronologies (here we plot all impacts of $D>$ 50 km objects) on the M-Dwarf-hosted, HZ planets TRAPPIST-1e, TRAPPIST-1f, TOI-700d and Proxima Centauri b.  In contrast to figure \ref{fig:impacts_plns}, this figure considers all collisional fragments to be volatile-rich (see text for additional discussion).  Time is plotted with respect to the end of the planet formation process (20 Myr).  The red dashed lines and grey shaded regions represent our determined limits for achieving an impact potentially capable of reconstituting an atmosphere on a desiccated planet: $D\geq$ 100 km and $t\geq$ 500 Myr.}
	\label{fig:impacts_frags}
\end{figure*}

To demonstrate a scenario where significant volatile-rich bombardment persists on $\sim$500 Myr timescales in our systems, we relax the requirement that impactors originate outside of the snowline when analyzing collisional fragment impacts in our systems.  While one might loosely justify this assumption by speculating that fragments might possess enhanced volatile contents by virtue of being ejected from near the surfaces of differentiated proto-planets, we caution the reader that this analysis is the least conservative of our study, and is meant to serve as more of a limiting case than a proof of concept.  Figure \ref{fig:impacts_frags} plots our derived fragment impact chronologies in the same manner as figure \ref{fig:impacts_plns}, and depicts appreciable bombardment approaching $\sim$Gyr timescales in our systems (particularly on the TRAPPIST planets).  It is important to note that, even under these favorable assumptions, large late impacts are not guaranteed to result in net-atmospheric enhancement.  Depending on the geometry of the impact \citep[e.g.:][]{svetsov07,shuvalov09}, such events can result in no volatile accretion, and even erode a significant fraction of the planets' existing atmosphere.  Thus, we conclude that, provided these systems do not host distant massive planets capable of dynamically perturbing debris onto HZ crossing orbits over lengthy timescales \citep[discussed further in the subsequent section:][]{quintana14_gp,dencs19}, the outlook for delayed, impact-driven atmospheric reconstitution of M-Dwarf-hosted planets is bleak.

\section{Discussion}
\label{sect:discuss}

Our numerical models of planet formation are admittedly simplified, particularly in comparison to state-of-the-art models of these processes in the solar system \citep[e.g.:][]{carter15,walsh19,woo21}.  Specifically, our simulations do not include a gas-disk treatment, utilize initial disk conditions that are largely contrived (though loosely based on the empirically determined minimum mass extrasolar nebula), and co-add remnant debris populations from a variety of realizations utilizing a range of initial conditions when studying late bombardment.   Nevertheless, it is worthwhile to briefly compare our findings with those of other similar investigations into the formation of habitable worlds around low-mass stars.  The first studies of such systems utilizing modern modeling algorithms include \citet{ray07_mdwarf}, \citet{lissauer07}, and \citet{ogihara09}.  An important difference between our work and these early studies is that, prior to the Kepler mission, modelers assumed that the mass of the solid component of planet forming disks scaled more or less linearly with stellar mass.  Thus, each author concluded that HZ planets formed in-situ around M-Dwarfs were likely much smaller than the Earth.  Disregarding discrepancies in planet mass, the final systems generated in our simulations are qualitatively similar to those from these past studies in terms of the number of planets formed and their radial separations.  \citet{ray07_mdwarf} and \citet{lissauer07} also argued that planets formed in-situ around M-Dwarfs are likely deficient in volatile elements as a result of radial evolution of the snowline's location \citep{kennedy08}.  Though we agree with this conclusion in principle, our simulations both with and without external giant planets demonstrate a high degree of radial mixing; resulting in the very efficient incorporation of distant disk planetesimals into HZ planets.  While this analysis presumes a constant location of the snowline, it is possible that outer disk planetesimals still obtain heightened volatile contents by accreting inward drifting pebbles from the cooler, icier regions of the disk. 

If disk mass indeed scales linearly with stellar mass \citep[which may not be a strict requirement given the measured occurrence rates of $M\gtrsim$ 1.0 $M_{\oplus}$ planets around such stars:][]{dressing15,gaidos17}, Earth-mass planets can still evolve into the HZ if they form further out in their star's natal disks before being implanted into the HZ via Type I migration \citep{ogihara09}.  More recently, \citet{miguel20} performed a robust statistical investigation of such a scenario, and found several systems that remarkably resembled the architectures of detected M-Dwarf-hosted exoplanets.  Moreover, the authors concluded that the proto-planet seeds of HZ planets forming in-situ likely possess significantly enhanced water-ice contents compared to those around solar-type stars.   

Irregardless of where the proto-planet seeds formed \citep[and the degree to which pebble accretion plays a role in their formation:][]{coleman19}, given the rapid accretion timescales in the HZs of low-mass stars, it is fairly inarguable that gas dynamics play a vital role in their formation.  While this is a notable shortcoming of our contemporary modeling work, it is difficult to estimate to what degree this affects our analysis of late bombardment in these systems.  In many instances, we observe fragment-generating giant impacts occurring at $t\gtrsim$ 15 Myr around the smallest stars in our simulation set.  Therefore, it is reasonable to argue that the final stages of our simulations (that produce fragments and strand planetesimals) provide a reasonable model of the late stages of both in-situ and migration-driven formation of HZ worlds.  Thus, we contend that our planet formation simulations are adequate for our purposes given their intended role of serving as inputs for our late bombardment computations.  Moreover, as these impacts transpire at timings in excess of the inferred accretion timescale of Mars \citep{Dauphas11,kruijer17_mars,costa20_mars} and some earlier estimates of the timing of the Moon-forming impact \citep[e.g.:][]{barboni17,thiemens19}, it is clear that the chronology of rocky planet accretion around low-mass stars is not necessarily exceptionally dissimilar from that of the solar system's rocky worlds.  

Another noteworthy limitation of our present study is related to our numerical implementation of fragmentation.  Specifically, it is difficult to ascertain whether our prescription accurately captures the actual physics at play in the aftermath of massive accretion events.  Hydrodynamical models typically conclude that the material ejected in massive, imperfect accretion events should predominantly be in the form of dust \citep{johnson12,watt21} that is easily re-acreated by the target body \citep{gladman09}.  Indeed, observed infrared emission from debris disks \citep[e.g.:][]{lisse09,weinberger11,rieke21} seemingly originating from massive embryo-scale impacts largely support dust being the primary collisional product \citep{genda15}.  As we simulate the remnant debris as massive, dynamically interacting particles \citep{lands12,chambers13}, it is reasonable to question whether or not such massive, $\sim$planetesimal-sized fragments indeed represent a significant fraction of the leftover debris population or not.  While it is clear that large break-up events were essential in transforming the primordial main belt's SFD, mapping these processes to the regime of large, differentiated bodies is challenging.  Nevertheless, iron-rich planets like Mercury and asteroids like Psyche strongly suggest that fragmenting collisions without re-accretion occurred in the solar system's distant past.  Thus, we argue that it is highly likely that collisional fragments must constitute some portion of the leftover debris population in both the solar system, and other systems of rocky terrestrial planets.  However, the total fragment-planetesimal mass ratio derived in our work is clearly an artifact of our numerical methodology, and a major source of uncertainty in our estimates of late bombardment in low-mass star systems (beyond the previously discussed variable of the remnant population's SFD).  Future detailed observation of the solar system's asteroid belt and high-resolution simulations of giant impacts and their aftermath will be crucial for constraining the relative contribution of collisional fragments in the flux of late impacts on HZ planets.
 
Perhaps the investigations most similar to our own in terms of focusing on late volatile delivery from small body reservoirs are \citet{dencs19} and \citet{ray21}.  While both sets of authors' primary focuses were the TRAPPIST-1 planets, their main conclusions were largely similar to our own.  Rather than deriving small body populations directly from planet formation simulations, \citet{dencs19} opted for an artificial set-up analogous to the solar system's Earth-Jupiter-Asteroid Belt system.  Specifically, the work considered a 5-50 $M_{\oplus}$ planet on various resonant and non-resonant orbits with the known seven planets.  As a result of perturbations from the additional planet, the flux of water-rich asteroids striking the HZ worlds was necessarily higher than observed in our simulations (upwards of a few percent of the total belt is accreted by TRAPPIST-1g and h over $\sim$kyr timescales in their models).  Conversely, \citet{ray21} utilized the constraint of resonant chain survival to place upper limits on late volatile delivery to the TRAPPIST-1 HZ planets, and considered cases where a rouge planetary embryo perturbs debris disk material, in addition to asteroid belt-analog setups similar to those in \citet{dencs19} and our current work.  The authors concluded that interactions with $\gtrsim$0.05-0.10 $M_{\oplus}$ worth of remnant planetesimals typically disrupt the resonant chain, and drive the system towards dynamical instability.  In this manner, we concur with both author's conclusions that the systems most likely to experience delayed volatile delivery capable of reconstituting secondary atmospheres would be those possessing exo-belts possibly implanted during the terrestrial planet formation process (particularly systems with giant planet companions) as demonstrated in our current manuscript.  Currently, their are no known $\sim$Earth-sized HZ planets around M-Dwarfs that also host longer-period giant planets.  Perhaps the most promising known example of such a configuration is GJ 229A; which hosts a Super-Earth in the HZ at 0.33 au and a Sub-Neptune at 0.89 au \citep{feng20a}.  However, a $\sim$Neptune-mass planet at $\sim$1.0  au is well within the range of possible undetected planets that might exist in the TRAPPIST-1 system as inferred from both transit timing variations \citep{jontof18} and legacy RV data \citep{boss17}.

An important mechanism omitted from the analyses of \citet{dencs19}, \citet{ray21}, and our own is size and semi-major axis dependent radial drift of small bodies due to anisotropic thermal emission \citep[the Yarkovsky effect, e.g.:][]{vokrouhlicky98,bottke01}.  Depending on the architecture of an exoplanet system, retrograde rotating debris particles of a particular size in an asteroid belt analog will drift inward at a particular semi-major axis decay rate ($da/dt$).  While this process is too efficient for extremely small objects on short-period orbits (i.e.: they are removed too quickly), appropriately distant objects with moderate sizes might represent a lucrative source of potential late bombardment.  Future work on late volatile delivery in systems of terrestrial worlds around low-mass stars should strive to comprehensively consider the potential of Yarkovsky drift to reconstitute atmospheres on desiccated planets.

\section{Conclusions}

In this paper, we presented a suite of numerical simulations designed to model in-situ rocky planet formation across the M-Dwarf mass spectrum.  From these initial models, we derived leftover fragment and planetesimal debris populations that might source appreciable delayed volatile delivery capable of reconstituting secondary atmospheres on desiccated planets.  With an additional batch of high-resolution models, we followed the dynamical evolution of the small bodies from these debris fields within three M-Dwarf-hosted, multi-planet systems (TRAPPIST-1, Proxima Centauri and TOI-700) containing $\sim$Earth-sized planets in the habitable zone (HZ).  Through this analysis we find that, even under the most favorable assumptions, debris is removed too quickly to represent a substantial volatile delivery mechanism after the host-stars' lengthy pre-Main Sequence cooling phase is complete (in excess of 1 Gyr for the smallest stars).  If we relax the requirement that debris originate outside of the snowline by considering all collisional fragments (ideally objects ejected from the surfaces of differentiated bodies) volatile-rich, we observe some $D\gtrsim$ 100 km impacts occurring at $t\gtrsim$ 500 Myr in our systems.  However, it is unclear whether the remnant fragment mass in these systems would be as high as assumed in our calculations given the fact that hydrodynamical models tend to predict that the majority of the mass liberated in giant embryo-embryo impacts should be in the form of dust that eventually falls back onto the progenitor object.

The major conclusion of our modeling work is that the most promising reservoir of small bodies capable of reconstituting atmospheres on M-Dwarf hosted HZ planets are exo-asteroid belts generated as a byproduct of the planet formation process.  While the presence of external ($P\sim$ 100-1,000 yr) giant planets ($\sim$Neptune-Saturn-mass) limits the efficiency of planetesimal implantation in such belts, their existence is crucial for perturbing objects onto HZ-crossing orbits \citep{dencs19}.  

While we do not plead to comprehensively explain rocky planet formation around M-Dwarfs with any level of sophistication given the simplistic setups employed in our numerical simulations, our derived impact chronologies should be viewed as a fairly model-independent demonstration of the non-viability of atmospheric reconstitution of desiccated planets via late bombardment.  We investigated a range of planet formation scenarios across the M-Dwarf mass spectrum (including models with and without giant planets, and scenarios that do not consider planetesimal distributions) and found that leftover debris fields and exo-asteroid belts are fairly ubiquitous relics of the process.  However, it is important to note that such small body reservoirs are extremely fragile during epochs of giant planet migration \citep[e.g.:][]{ray09b,ray10,minton10,clement18_ab}.  As high-precession RV surveys and forthcoming mircolensing monitoring by the Roman Space Telescope \citep{johnson20} continue to discover longer-period giant planets around M-Dwarfs \citep[e.g.:][]{tuomi14,feng20a,feng20b}, dynamical models must begin to investigate whether their orbital architectures evidence past epochs of violent instability or large-scale migration capable of destroying terrestrial planets and significantly eroding small body reservoirs.

\section*{Acknowledgments}

We thank Tim Lichtenberg, John Chambers and an anonymous reviewer for useful comments and insight that greatly improved the manuscript and the presentation of the results. The work described in this paper was supported by Carnegie Science's Scientific Computing Committee for High-Performance Computing (hpc.carnegiescience.edu).  This work also used the Extreme Science and Engineering Discovery Environment (XSEDE), which is supported by National Science Foundation grant number ACI-1548562. Specifically, it used the Comet system at the San Diego Supercomputing Center (SDSC).  This research was done using resources provided by the Open Science Grid \citep{osg1,osg2}, which is supported by the National Science Foundation award 1148698, and the U.S. Department of Energy's Office of Science.  Some of the computing for this project was performed at the OU Supercomputing Center for Education and Research (OSCER) at the University of Oklahoma (OU). This research was supported in part through research cyberinfrastructure resources and services provided by the Partnership for an Advanced Computing Environment (PACE) at the Georgia Institute of Technology.

\bibliographystyle{apj}
\newcommand{\sci}{$Science$ }
\newcommand{\psj}{$PSJ$ }
\bibliography{mdwarf.bib}
\end{document}